\documentclass[a4paper, fleqn, onecolumn]{article} 

\usepackage[margin=2cm]{geometry} 
\usepackage{makecell}
\usepackage{longtable}
\usepackage{adjustbox}
\usepackage{rotating}
\usepackage{lscape}
\usepackage{xcolor}
\usepackage{url}
\usepackage{booktabs}
\usepackage{tabularx} 
\usepackage{fancyhdr}
\usepackage{lastpage}

\begin{document}

\title{Machine Learning on Blockchain Data: A Systematic Mapping Study}

\author{
Georgios Palaiokrassas\textsuperscript{1,2}, Sarah Bouraga\textsuperscript{3}, Leandros Tassiulas\textsuperscript{1,2}\\\\
\textsuperscript{1}Yale Institute for Network Science, Yale University, USA\\
\textsuperscript{2}Department of Electrical Engineering, Yale University, USA\\
\textsuperscript{3}Namur Digital Institute (NADI), Belgium\\
\texttt{\small \{georgios.palaiokrassas, leandros.tassiulas\}@yale.edu},\\
\texttt{\small sarah.bouraga@unamur.be}
}

\date{} 

\maketitle

\begin{abstract}
\textbf{Context}: Blockchain technology has drawn growing attention in the literature and in practice. Blockchain technology generates considerable amounts of data and has thus been a topic of interest for Machine Learning (ML). 

\textbf{Objective}: The objective of this paper is to provide a comprehensive review of the state of the art on machine learning applied to blockchain data. This work aims to systematically identify, analyze, and classify the literature on ML applied to blockchain data. This will allow us to discover the fields where more effort should be placed in future research. 

\textbf{Method}: A systematic mapping study has been conducted to identify the relevant literature. Ultimately, 159 articles were selected and classified according to various dimensions, specifically, the domain use case, the blockchain, the data, and the machine learning models. 

\textbf{Results}: The majority of the papers (49.7\%) fall within the Anomaly use case. Bitcoin  (47.2\%) was the blockchain that drew the most attention. A dataset consisting of more than 1.000.000 data points was used by 31.4\% of the papers. And Classification (46.5\%) was the ML task most applied to blockchain data. 

\textbf{Conclusion}: The results confirm that ML applied to blockchain data is a relevant and a growing topic of interest both in the literature and in practice. Nevertheless, some open challenges and gaps remain, which can lead to future research directions. Specifically, we identify novel machine learning algorithms, the lack of a standardization framework, blockchain scalability issues and cross-chain interactions as areas worth exploring in the future.  

\vspace{0.5cm} 

\end{abstract}

\maketitle

\section{Introduction}
Blockchain technology has sparked a lot of interest over the years. A particularly interesting characteristic of the technology is the transparency it offers. Indeed, all the transactions recorded on a public blockchain (amounting to hundreds of thousands of transactions a day, just for Bitcoin) can be viewed, retrieved and analyzed by anyone. This is a huge paradigm shift, compared to incumbent institutions such as traditional banks. 

The amount of available blockchain data offers a lot of potential for analysis. We can analyze the blockchain data to discover unknown patterns in the data, or to predict the next cryptocurrency price, or to detect fraud to name a few. In order to carry out these types of analyses, we can use machine learning, a subfield of Artificial Intelligence. Since machine learning requires a lot of data to perform well, and since blockchain data are public and available in large quantities, this seems like a match made in heaven. 

This claim is supported by the plethora of scientific articles we analyze here. Many researchers addressed the questions raised above, i.e. they tried to discover hidden patterns in blockchain data, they proposed solutions for the prediction of cryptocurrency prices, others focused on the detection of fraudulent activity on a blockchain, and multiple other use cases. 

Due to the rapid evolution of both technologies (blockchain and machine learning), it is not trivial to keep track of the state of the art: What has been done? How? On which platform?... We believe it is essential for researchers and practitioners to have a clear view of the current state of the art. On the one hand, practitioners need to know the new solutions for a given problem, taking advantage of the latest advances in a given field. Researchers, on the other hand, need to know the works they can build on and the research directions they can pursue. 

Hence, the aim and corresponding contribution of this paper is to propose a \textbf{Systematic Mapping Study} of scientific papers applying machine learning to blockchain data. As explained in Section \ref{sec:methodology}, we apply the rigorous methodology and follow the guidelines proposed by \cite{keele2007guidelines,petersen2008systematic}. Specifically, our contribution is fourfold: 
\begin{enumerate}
    \item We \textbf{identify, analyze and organize 159 papers}, published between 2008 and 2023, applying machine learning to blockchain data. Section \ref{sec:studies} shows the distribution of the papers across venues (Book chapter, Conference proceedings, and journal articles) and their evolution throughout the years. 
    \item Using keywording, we propose a \textbf{classification scheme} organizing the studies across multiple dimensions, namely: the Use Case, the Blockchain, the Data, and the Machine Learning task. We present the classification scheme in Section \ref{sec:classification-scheme}. 
    \item We propose a \textbf{mapping of the studies}, focusing on the use cases: Address classification, Anomaly detection, Cryptocurrency price prediction, Performance prediction, and Smart contract vulnerability detection. We present the mapping results in Section \ref{sec:mapping-results}. 
    \item We identify potential \textbf{research gaps} in Section \ref{sec:discussion}, where we also discuss the results of this work. 
\end{enumerate}

The main aim of this work is to investigate the current state of the art regarding machine learning on blockchain data. Specifically, we aim to assess the academic publication aspect, such as the popular forums and publication types; as well as the technical aspects, such as the blockchains that have been analyzed, the datasets used and/or curated by researchers, and the machine learning algorithms that were applied.
We hope the study will provide the reader with a clear overview of the current research of machine learning on blockchain data, and that it will highlight potential research gaps, suggesting potential future research directions for researchers. 

\subsection{Related Surveys}
Various authors proposed related surveys, which we classified into three categories: (i) general surveys about blockchain and machine learning, (ii) surveys about blockchain and machine learning in a specific sector/industry or application, and (iii) systematic reviews. 

Some researchers proposed an overview of blockchain technology and of artificial intelligence applications in the blockchain. Siddiqui \& Haroon \cite{siddiqui2022application} introduced blockchain technology, its characteristics and benefits; then explored how AI and blockchain can be combined in order to mitigate their limitations. The authors also offered a wide range of examples of applications of these two technologies; and discussed the case of edge computing. Similarly, Inbaraj \& Chaitanya \cite{inbaraj2020need} provided an overview of blockchain and AI, discussed the implementation of blockchain in various industries, and discussed the integration of blockchain and AI in various applications. 

Other works focused on blockchain and artificial intelligence in a specific sector or for a specific application. A number of authors proposed an analysis of existing works addressing blockchain and machine learning related to the Internet of Things (IoT). Aoun et al. \cite{aoun2021review} proposed an overview of IoT, its challenges and how blockchain technology can help in the development of the Industry 4.0. 
In \cite{wu2021deep}, the authors provided a summary and analysis of such works using three perspectives, namely: consensus mechanism, storage, and communication. The authors discussed how blockchain and machine learning interact in Industrial IoT (IIoT) and highlight the security and privacy risks of such solutions. Liu et al. \cite{liu2022survey} analyzed works exploring blockchain-enabled federated learning in the context of Digital Twin, from aspects pertaining to security, fault-tolerance, fairness, efficiency, cost-saving, profitability, and support for heterogeneity. Various authors addressed the issue of IoT security. On the one hand, Williams et al. \cite{williams2022survey} reviewed the security in IoT and emphasized the impact of emerging technologies, including fog/edge/cloud computing and quantum computing in addition to blockchain and machine learning. On the other hand, in \cite{mohanta2020survey}, the authors analyzed how machine learning, artificial intelligence, and blockchain technology can help to address IoT security. In \cite{hua2022applications}, the authors discussed how we can integrate blockchain technology and artificial intelligence with smart grids in order to facilitate prosumers' participation in energy markets. 

Han et al. \cite{han2023accounting} adopted the same approach but applied it to the case of accounting and auditing. Cheng et al. \cite{cheng2021integration} reported on the use of machine learning and blockchain technology in the healthcare sector, and more specifically discussed the implications for cancer care. 

A study closer to our research here would be \cite{ren2022past}, where the authors analyzed works related to cryptocurrency research and machine learning. The findings of this work: (i) confirm the popularity of the topic in research; (ii) show that cryptocurrency price (or related) prediction is the most popular topic; and (iii) indicate that many different algorithms are used and that common problems such as overfitting and interpretability are still present. 

Mainly two elements differentiate our work here from these related surveys. Firstly, we do not focus on a specific area, but provide an overview of existing works pertaining to a wide range of industries, sectors, and applications. In particular, we do not restrict our analysis to cryptocurrency as in \cite{ren2022past} but our review encompasses works related to a wide range of use cases. Secondly, we focus on machine learning on blockchain data; while many of the mentioned surveys addressed works about the integration of the two technologies to a specific problem or opportunity. 

Finally, we identified a bibliometric analysis, two systematic reviews, and a survey. Bai \& Sarkis \cite{bai2022critical} conducted a bibliometric and network analysis to identify research areas related to formal analytical modeling for blockchain in the domain of supply chains. In \cite{miglani2021blockchain}, the authors performed a systematic review of the use of blockchain management and machine learning in the particular context of IoT environment. Lin et al. \cite{lin2022systematic} carried out a systematic review and analyzed 25 papers treating unsupervised learning, supervised learning and topological analysis for the detection of illicit transactions on the Bitcoin blockchain. Finally, a comprehensive survey was conducted in \cite{hassan2022anomaly}, where the authors depicted the state of the art regarding anomaly detection in blockchain networks. They analyzed papers and classified them based on the blockchain layers, namely, the data layer, the network layer, the incentive layer, and the smart contract layer. 

We believe our work is different from the mentioned surveys and hence adds value to the literature. Indeed, the scope we consider here is broader than the ones addressed in \cite{bai2022critical,miglani2021blockchain,lin2022systematic,hassan2022anomaly}. We propose a coarse-grained analysis of the state of the art regarding machine learning applied to blockchain data. We do not focus here on a particular use case or industry. In addition, conducting our review now allows us to consider recently published papers that could not have been treated by the previous surveys. Finally, for the coarse-grained analysis, we propose different classification dimensions than the ones discussed in previous surveys. The current work complements thus the existing surveys proposed by fellow researchers.

\section{Background}
As stated above, the availability of large volumes of blockchain data offers a great opportunity for analysis. Classical machine learning tasks - such as hidden patterns, trends, or outliers detection - can be performed on these data. In the following subsections, we offer a background on blockchain technology and on machine learning.

\subsection{Blockchain}
Blockchain was originally proposed in 2008 as the accounting method for Bitcoin cryptocurrency \cite{nakamoto2008bitcoin}. 
The technology and ideas evolved in the years that followed and many blockchains and altcoins were introduced. A milestone for the course of blockchain technology was the introduction of Ethereum in 2015 \cite{buterin2014next}, an open decentralized blockchain platform which provides a virtual computing environment called Ethereum Virtual Machine, but also a Turing complete programming language to write smart contracts. Ethereum smart contracts are actually program instances running on the decentralized network and any user is allowed to deploy smart contracts enabling the development of different Decentralized Applications (DApps) with potential for different fields such as IoT \cite{palaiokrassas2021combining}, copyright management \cite{palaiokrassas2019deploying}, supply chain management \cite{queiroz2020blockchain}, healthcare \cite{agbo2019blockchain}, energy \cite{andoni2019blockchain}, Decentralized Finance (DeFi) \cite{werner2021sok} and many more.

Blockchain is actually a distributed ledger or distributed database recording digital transactions between two parties without the need for Third Trusted Parties (TTP). This allows the interaction of users without an intermediary, while anonymity is preserved, as one of the key blockchain's features. Every participating node in this peer-to-peer network has an actual "copy" of all the transactions that took place, ensuring their immutability and enhancing security. Security is also preserved by the utilization of cryptography, which is one of the principal aspects of blockchain technology.  

The transactions within a ledger are verified by multiple nodes or “validators,” within the cryptocurrency's peer-to-peer network using one of many varied consensus algorithms for resolving the problem of reliability in a network involving multiple unreliable nodes. 
Different consensus mechanisms have been proposed allowing the network of nodes to agree on the state of a blockchain. 
The most widely used consensus algorithms are the Proof of Work (PoW) algorithm and the Proof of Stake (PoS) algorithm  \cite{bach2018comparative}.  Bitcoin for example uses a PoW-based consensus protocol, while Ethereum transitioned from a PoW to PoS-based consensus protocol.
However, there are also other consensus algorithms, which utilize alternative implementations of PoW and PoS, as well as other hybrid implementations and some altogether new consensus strategies.

An increasing amount of blockchain data is generated through various interactions and activities. The type of data varies depending on the type of blockchains, the supported functionalities and remains unchanged given the immutable nature of ledgers. 
Blockchain data consists mainly of transaction and block data, smart contract logs, events, and interactions, network activity, and topology, while data analysis and machine learning techniques have been applied to them combining often external datasets such as cryptocurrency prices, news feed and public sentiment.

\subsection{Machine Learning}
As stated by M. Jordan and T. Mitchell, Machine Learning addresses the question of how to build computers that improve automatically through experience. It is one of today’s most rapidly growing technical fields, lying at the intersection of computer science and statistics, and at the core of artificial intelligence and data science \cite{jordan2015machine}. According to another definition by Deisenroth et al. \cite{deisenroth2020mathematics}, Machine learning is about designing algorithms that automatically extract valuable information from data with emphasis on “automatic”, i.e., machine learning is concerned about general-purpose methodologies that can be applied to many datasets, while producing something that is meaningful. 

A learning problem can be defined as the problem of improving some measure of performance when executing some task, through some type of training experience.
For example, in our study we identified several works attempting to learn to detect fraud in blockchain transactions, where the task is to assign a label of “fraud” or “not fraud” to any given
blockchain transaction. The performance metric
to be improved might be the accuracy of this
fraud classifier, and the training experience might consist of a collection of historical transactions, each labeled in retrospect as fraudulent or not.

A typical workflow of an ML framework commonly consists of the training phase and the testing phase, while sometimes the validation phase is part of the flow \cite{liu2020blockchain}. It could be noted that other steps are involved in some other categories of tasks and methods such as Reinforcement Learning or Federated Learning. 
In general, ML techniques can be classified into four different areas:

 \textbf{(i) Supervised learning} where the learning algorithm learns from labeled data. The training data take the form of a collection of (x, y) pairs and the goal is to produce a prediction y* in response to a query x*, 

\textbf{(ii) Unsupervised learning} generally involves the analysis of unlabeled data under assumptions about structural properties of the data.

 \textbf{(iii) Semi-supervised learning} is the branch of machine learning concerned with using labeled as well as unlabeled data to perform certain learning tasks. Conceptually situated between supervised and unsupervised learning, it permits harnessing the large amounts of unlabeled data available in many use cases in combination with typically smaller sets of labeled data \cite{van2020survey}. 

 \textbf{(iv) Reinforcement learning} where the information
available in the training data is intermediate
between supervised and unsupervised
learning. Instead of training examples that indicate
the correct output for a given input, the
training data in reinforcement learning are assumed
to provide only an indication as to whether
an action is correct or not.

Regardless of the area or the task, there are some steps appearing in the workflows of the different ML solutions and research efforts. The most common ones are: i) Data Collection; (ii) Data Preprocessing; (iii) Feature Extraction and (iv) Algorithm Selection.

\section{Methodology} \label{sec:methodology}
Following \cite{abdelmaboud2015quality}, we conducted this systematic mapping study by considering both the guidelines proposed by \cite{petersen2008systematic} and by \cite{keele2007guidelines}. In this section, we detail the steps we performed in order to carry out our study.

\subsection{The Research Questions}
Systematic mapping studies, in general, aim to provide an overview of the current research in a given area \cite{petersen2008systematic}. In this study, we focus on blockchain and machine learning; and we offer a coarse-grained analysis of the current research, we identify the quantity and type of research, and the gaps and future research directions. Specifically, the overarching goal can be defined by the following research questions: 

\begin{itemize}
    \item \textbf{RQ1.} Which topics related to machine learning on blockchain have been investigated and to what extent?
    \item \textbf{RQ2.} What diverse types of blockchain data have been analyzed and to what extent has each type been represented? 
    \item \textbf{RQ3.} Which machine learning types of models have been applied on blockchain data and to what extent has each type of model been represented? 
    \item \textbf{RQ4.} In which forums has research on machine learning on blockchain  been published?
\end{itemize}

Answering these research questions will give us a comprehensive overview of the current state of research, and thereby addressing the main goal of this paper.

\subsection{Data Sources and Search Strategy}
The database sources that have been used as primary sources in this study are the following: 
\begin{itemize}
    \item \textbf{Google Scholar}: \url{https://scholar.google.com}
    \item \textbf{Springer}: \url{https://link.springer.com}
    \item \textbf{ScienceDirect}: \url{https://www.sciencedirect.com}
\end{itemize}
These are some of the most commonly used database sources in software engineering. 
We started the study in December 2022 and, given the relative novelty of the technology, we decided to search publications without any period limitation. 

The first step was to define search keywords as recommended by \cite{keele2007guidelines}. We considered the terms ``Blockchain", ``Smart contract", ``Decentralized application", ``Bitcoin", ``Ethereum", ``Machine learning", ``Analytics", and ``Artificial intelligence" as our keywords. We used the logical operators OR and AND to link the main keywords. For some database sources, we conducted several searches to cover all keywords. Specifically, we started for instance with (Blockchain AND ``Machine learning"), retrieved the returned publications, and repeated the search with the other keywords combinations. For other database sources, we used the following single search string: 
\begin{itemize}
    \item ALL((blockchain or “DApp” or "Ethereum" or “Bitcoin” or “smart contract” or "decentralized application") AND (“machine learning” or “analytics” or "artificial intelligence" or "ai"))
\end{itemize}

The search strings were applied to search the database sources by considering the title, abstract and keywords.

\subsection{Study Selection}
We developed inclusion and exclusion criteria in order to select the most relevant and important publications. We confronted the title, abstract and full text of the articles previously retrieved, in order to ensure that the publications fit the scope of our study. 

We excluded studies that exclusively provided a discussion or a conceptual solution of blockchain and machine learning (as in \cite{tsai2021efficient,wang2019analytic,kalafatelis2021island,pal2022blockchain}). The reason for this exclusion is that we are interested in the data and models used to address a specific problem. Secondly, we are interested in the application of machine learning on blockchain data. Hence, the papers addressing the use of blockchain technology to enhance the performance of machine learning were not included (such as \cite{ouyang2022intelligent,YOU2022Curvetime}). Similarly, publications analyzing data without any features pertaining to blockchain data were excluded, for instance cryptocurrency price prediction that only use data about the historical prices (as in \cite{atsalakis2019bitcoin,lahmiri2020intelligent,manahov2021efficiency,ciaian2018virtual,lahmiri2020big,liu2021bitcoin,rathore2022real}) or that only use traffic data \cite{vesely2019detect,sundareswaran2022packet}. Furthermore, articles where no machine learning model was applied to the data were also excluded, for instance articles providing a descriptive statistics analysis (such as \cite{zheng2021xblock,kanth2022parameter}). Finally, surveys were also excluded since they do not provide a new machine learning solution but discuss and analyze existing ones \cite{ren2022past,siddiqui2022application,inbaraj2020need}. 

To summarize, the inclusion and exclusion criteria applied in this study are the following. 

The inclusion criteria: 
\begin{itemize}
    \item Articles must report on the application of machine learning to blockchain data
    \item Articles must be peer-reviewed 
    \item Articles must be written in English
    \item Articles are published in or after 2008
    \item Articles must be published
\end{itemize}

The exclusion criteria: 
\begin{itemize}
    \item Articles propose a discussion or a conceptual solution of blockchain and machine learning, such as surveys
    \item Articles are not written in English 
    \item Articles are not peer-reviewed 
    \item Articles are published before 2008
    \item Articles are on pre-press
\end{itemize}

The selection of the primary studies was done in four phases and was conducted by two reviewers who examined all the retrieved studies. For each study, there was a data extractor. The data extraction form was then checked by the other reviewer. If any conflicts were raised, the authors resolved them by examining the paper together.  The four phases of the selection are the following:
\begin{enumerate}
    \item \textbf{Phase 0. Application of search strings to database sources} \textit{- Number of papers included in the next phase: 3434.} The total number of articles retrieved from the database sources (Springer, ScienceDirect and Google Scholar) was 3434. The results of these articles were included in the next phase. 
    \item \textbf{Phase 1. Title-based selection.} \textit{- Number of papers included in the next phase: 376.} In this phase, we read and assessed the paper's title using our inclusion and exclusion criteria. If the paper fit the scope of the study, it was included in the next phase. Otherwise, it was discarded. When the evaluation of the title was difficult and/or led to debate, we decided to include the paper in the next phase.     
    \item \textbf{Phase 2. Removal of duplicates.} \textit{- Number of papers included in the next phase: 304.} We found 72 duplicates, we removed them from our data and we included the remaining 304 papers in the next phase. 
    \item \textbf{Phase 3. Abstract-based selection.} \textit{- Number of papers included in the next phase: 131.} We read the abstract and keywords of each paper. It allowed us to confirm or infirm their selection and inclusion in the next phase. We discarded 173 papers in this phase and included the remaining 131 in the next phase. 
    \item \textbf{Phase 4. Full text selection.} \textit{- Number of papers included in the next phase: 85.} In this phase, we read the full text of the 131 selected papers from phase 3. This allowed us to make the final selection and make sure that the articles included in the study were actually related to the study. Using our inclusion and exclusion criteria, we further discarded 46 articles. This resulted in a final set of primary studies consisting of 86 articles. We recorded the basic information of each of these papers in an Excel sheet, namely: the title, the authors, the publication type and the year of publication. 
    \item \textbf{Phase 5. Secondary search.} \textit{- Number of papers included: 159.} In this phase, we conducted an additional selection of papers based on the references of the articles included in Phase 4. This allowed us to include papers that we may have missed in earlier phases. We proceeded in the same way as above. Specifically, we first selected the papers based on the title, and we removed the duplicates. We then read the abstract, the keywords, and the full text. In order to be included in the final set of studies, the papers had to satisfy the inclusion and exclusion criteria. This phase allowed us to include 74 additional papers, which resulted in a final set of studies consisting of 159 articles. 
\end{enumerate}

\begin{figure}
	\centering
		\includegraphics[scale=0.4]{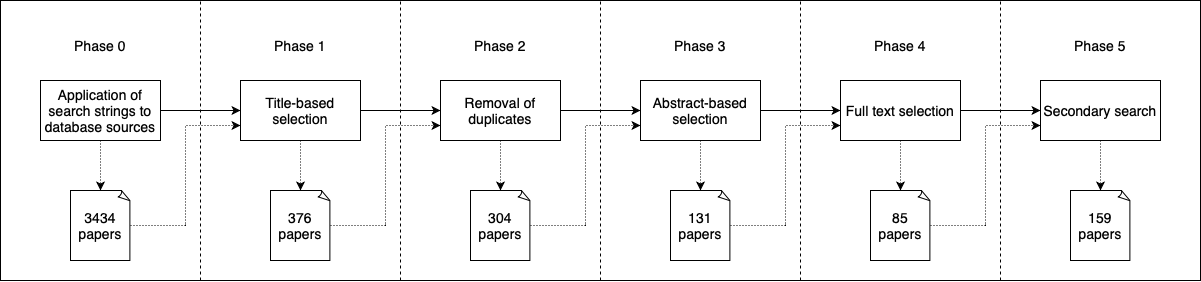}
	\caption{Selection Process}
	\label{fig:selection-process}
\end{figure}

This process is summarized in Figure \ref{fig:selection-process}. 

\section{Studies} \label{sec:studies}
The studies selected in this systematic mapping study consisted of 159 articles taking the form of journal articles, conference proceedings, book chapters and workshop proceedings from 2015 to 2023. In order to address RQ4, we identified the publication years and forums of the studies. Tables \ref{tab:forums} and \ref{tab:distribution-publication-type} show a summary of the publication forums of the studies, and the distribution of the studies by publication type respectively. We can see that the \textit{IEEE Access}, \textit{Expert Systems with Applications}, and the \textit{International Conference on Blockchain and Trustworthy Systems} are the most popular forums; and that 2022 and 2021 were the years where most research publications were proposed.

\begin{longtable}{p{13cm}p{2.5cm}}
\toprule
 Publication Source & Count of papers  \\ 
\midrule
\textit{Book chapter}           & 1 \\
Blockchain Intelligence         & 1 \\
\textit{Conference proceedings} & 82  \\
International Conference on Blockchain and Trustworthy Systems & 7 \\
International Conference on Network and System Security & 3  \\
Hawaii International Conference on System Sciences & 2 \\
IEEE International Conference on Blockchain & 2 \\
IEEE International Conference on Blockchain and Cryptocurrency (ICBC) & 2 \\
IEEE International Symposium on Circuits and Systems (ISCAS) & 2 \\
International Conference on Complex Networks and Their Applications  & 2 \\
ACM International Conference on AI in Finance (ICAIF) & 1 \\
ACSAC: Annual Computer Security Applications Conference & 1 \\
Advances in Knowledge Discovery and Data Mining & 1 \\
Biometric and surveillance technology for human and activity identification XII & 1 \\
Blockchain and Trustworthy Systems & 1 \\
Blockchain Research and Applications for Innovative Networks and Services (BRAINS) & 1 \\
CAAI International Conference on Artificial Intelligence & 1 \\
Crypto Valley Conference on Blockchain Technology (CVCBT) & 1 \\
IEEE Annual Computers, Software, and Applications Conference (COMPSAC) & 1 \\
IEEE Confs on Internet of Things, Green Computing and Communications, Cyber, Physical and Social 
IEEE Global Communications Conference & 1 \\
IEEE Global Engineering Education Conference  & 1 \\
IEEE Int Conf on High Performance Computing \& Communications; Int Conf on Data Science \& Systems; Int Conf on Smart City; Int Conf on Dependability in Sensor, Cloud \& Big Data Systems \& Application (HPCC/DSS/SmartCity/DependSys) & 1 \\
IEEE International Conference on Big Data (Big Data) & 1 \\
IEEE International Conference on Decentralized Applications and Infrastructures (DAPPS) & 1 \\
IEEE International Conference on Machine Learning and Applications & 1 \\
IEEE International Conference on Software Quality, Reliability and Security & 1 \\
IEEE Symposium Series on Computational Intelligence & 1 \\
Information Science and Applications & 1 \\
Information Security for South Africa & 1 \\
Information Systems in a Changing Economy and Society & 1 \\
Intelligent Computing and Innovation on Data Science & 1 \\
International AAAI Conference on Web and Social Media  & 1 \\
International Conference of Smart Systems and Emerging Technologies (SMARTTECH) & 1 \\
International Conference on Advanced Communication Technology & 1 \\
International Conference on Advanced Data Mining and Applications & 1 \\
International Conference on Advances in Cyber Security & 1 \\
International Conference on Algorithms and Architectures for Parallel Processing & 1 \\
International Conference on Automated Software Engineering & 1 \\
International Conference on Big Data, Information and Computer Network (BDICN) & 1 \\
International Conference on Big Data and Security & 1 \\
International Conference on Blockchain Computing and Applications (BCCA) & 1 \\
International Conference on Computer Communication and Networks & 1 \\
International Conference on Computer Theory and Applications (ICCTA) & 1 \\
International Conference on Data Mining Workshops & 1 \\
International Conference on Data Science and Computer Application (ICDSCA) & 1 \\
International Conference on Deep Learning, Big Data and Blockchain & 1 \\
International Conference on Digital Forensics & 1 \\
International Conference on Information Security Theory and Practice & 1 \\
International Conference on Information Technology (ICIT) & 1 \\
International Conference on Information Technology and Quantitative Management & 1 \\
International Conference on Intelligence and Security Informatics & 1 \\
International Conference on Internet of Things: Systems, Management and Security & 1 \\
International Conference on Internet Measurement & 1 \\
International Conference on Machine Learning Technologies (ICMLT) & 1 \\
International Conference on Mobile Ad Hoc and Smart Systems (MASS) & 1 \\
International Conference on Mobile Networks and Management & 1 \\
International Conference on Parallel, Distributed, and Network-Based Processing & 1 \\
International Conference on Privacy, Security and Trust & 1 \\
International Conference on Service-Oriented System Engineering & 1 \\
International Conference on Smart City Applications & 1 \\
International Conference on Soft Computing Models in Industrial and Environmental Applications (SOCO) & 1 \\
International Conference on Trust, Security and Privacy in Computing and Communications (TrustCom) & 1 \\
International Conference on Ubiquitous and Future Networks (ICUFN) & 1 \\
International Conference Web Information Systems Engineering & 1 \\
International Congress on Blockchain and Applications & 1 \\
International Joint Conference on Neural Networks & 1 \\
International Symposium on Software Reliability Engineering & 1 \\
Italian Conference on CyberSecurity & 1 \\
Pacific-Asia Conference on Advances in Knowledge Discovery and Data Mining & 1 \\
Recent Trends in Analysis of Images, Social Networks and Texts & 1 \\
World Wide Web Conference & 1 \\
\textit{Journal}                &  73 \\
IEEE Access & 5 \\
Expert Systems with Applications & 4 \\
MDPI Sensors  & 3 \\
Applied Soft Computing & 2 \\
Blockchain: Research and Applications & 2 \\
IEEE Transactions on Computational Social Systems & 2 \\
IEEE Transactions on Systems, Man, and Cybernetics: Systems & 2 \\
IEEE Transactions on Network Science and Engineering & 2 \\
Information Processing and Management & 2 \\
Mathematics & 2 \\
MDPI Electronics & 2 \\
PloS one & 2 \\
Academy of Accounting and Financial Studies Journal & 1 \\
ACM SIGMETRICS Performance Evaluation Review & 1 \\
ACM Transactions on Internet Technology & 1 \\
Applied Intelligence & 1 \\
CCF Transactions on Pervasive Computing and Interaction & 1 \\
Computational Economics & 1 \\
Computer Communications & 1 \\
Computer Networks & 1 \\
Data \& Knowledge Engineering & 1 \\
Data Science and Management  & 1 \\
Decision Support Systems & 1 \\
EPJ Data Science & 1 \\
Eurasian Economic Review & 1 \\
Finance Research Letters & 1 \\
Future Generation Computer Systems & 1 \\
IEEE Systems Journal & 1 \\
IEEE Transactions on Circuits and Systems II: Express Briefs & 1 \\
IEEE Transactions on Dependable and Secure Computing  & 1 \\
IEEE Transactions on Information Forensics and Security & 1 \\
International Journal of Information Technology  & 1 \\
International Journal of Forecasting & 1 \\
International Journal of Information Security & 1 \\
IOP SciNotes & 1 \\
Journal of Behavioral and Experimental Finance & 1 \\
Journal of Combinatorial Optimization  & 1 \\
Journal of Computational and Applied Mathematics & 1 \\
Journal of Finance and Data Science & 1 \\
Journal of Management Information Systems & 1 \\
Journal of Risk and Financial Management & 1 \\
Journal of Supercomputing & 1 \\
Journal of Systems Architecture & 1 \\
Knowledge-Based Systems & 1 \\
Measurement: Sensors & 1 \\
Multimedia Tools and Applications & 1 \\
Neural computing and applications & 1 \\
Neural Processing Letters & 1 \\
North American Journal of Economics and Finance & 1 \\
Pattern Recognition Letters & 1 \\
Peer-to-Peer Networking and Applications & 1 \\
Physica A & 1 \\
Science China Information Sciences & 1 \\
Security and Communication Networks & 1 \\
SN Computer Science & 1 \\

\textit{Workshop proceedings}   &  3 \\

International Workshop on Emerging Trends in Software Engineering for Blockchain & 1 \\
International Workhsop on Financial Cryptography and Data Security & 1 \\
International Workshops of ECML PKDD on Machine Learning and Principles and Practice of Knowledge Discovery in Databases & 1 \\
Total & 159 \\
\bottomrule
\caption{Publication Forums of Primary Studies}
\end{longtable}
\label{tab:forums}

\begin{table}[htbp]
\caption{Distribution of Studies by Publication Type.}\label{tab:distribution-publication-type}
\begin{tabularx}{\textwidth}{l*{12}{>{\raggedright\arraybackslash}X}}
\toprule
\textbf{Publication Type} & \textbf{2013} & \textbf{2014} & \textbf{2015} & \textbf{2016} & \textbf{2017} & \textbf{2018} & \textbf{2019} & \textbf{2020} & \textbf{2021} & \textbf{2022} & \textbf{2023} & \textbf{Total} \\
\midrule
\textbf{Book Chapter} & 0 & 0 & 0 & 0 & 0 & 0 & 0 & 0 & 1 & 0 & 0 & 1\\
\textbf{Conference Proceedings} & 1 & 0 & 4 & 2 & 3 & 12 & 14 & 17 & 19 & 9 & 1 & 82\\
\textbf{Journal} & 0 & 0 & 1 & 0 & 2 & 3 & 6 & 13 & 16 & 26 & 6 & 73\\
\textbf{Workshop Proceedings} & 0 & 0 & 0 & 0 & 1 & 0 & 1 & 0 & 0 & 1 & 0 & 3\\
\hline
\textbf{Total} & 1 & 0 & 5 & 2 & 6 & 15 & 21 & 30 & 36 & 36 & 7 & 159\\
\bottomrule
\end{tabularx}
\end{table}

\section{Classification Scheme and Distribution Trend} \label{sec:classification-scheme}
Following \cite{petersen2008systematic}, we designed our classification scheme by keywording. Specifically, we identified relevant keywords by reading abstract and this set of keywords allowed us to define a number of dimensions that we grouped into high level categories in order to form a structured framework. We should note that the list of keywords evolved with our reading, i.e. new keywords were added to the list as we analyzed new publications. 

The studies were classified from four perspectives: (i) use case, (ii) blockchain, (iii) data, and (iv) machine learning. 
We should note that it is common to classify studies in software engineering based on the contribution type and the research type \cite{petersen2008systematic}. However, given the focus of this study, we elected not to use these dimensions. Indeed, we selected papers applying machine learning to blockchain data in order to fulfill a particular goal. Hence, most studies provide a ``Method" contribution (Contribution type) and propose a ``Solution" (Research type). We believe that incorporating these dimensions in our analysis would not have given the reader a lot of valuable insight. 

With the classification scheme clearly defined, we sorted the relevant articles into this scheme. We used an Excel table to document the data extraction process. The table contained each category of the classification scheme. As mentioned earlier, only one reviewer filled the data extraction form (i.e. a row in our Excel table) and the other reviewer checked the form afterwards. Conflicts were resolved by analyzing and discussing the paper together. 
Once all conflicts were resolved, we computed the frequencies of publications for each category.

\subsection{Use Case and Distribution}
Using the keywording technique described in \cite{petersen2008systematic}, we identified five major use cases: 
\begin{itemize}
    \item \textbf{Address Classification}. While pseudo-anonymity is a major property of blockchains, studies have addressed the de-anonymization of blockchain users. This use case pertains to user identification, address classification, address clustering, etc. 
    \item \textbf{Anomaly Detection}. This use case relates to the detection of any suspicious behavior on a blockchain, e.g. Ponzi Scheme detection, Illicit transaction detection, Attack detection, etc. 
    \item \textbf{Cryptocurrency Price Prediction}. This use case pertains to the prediction of cryptocurrency price, e.g. Bitcoin, Ether, etc. 
    \item \textbf{Performance Prediction}. Scalability is an important challenge for the blockchain community. This use case pertains to the prediction of performance, such as the prediction of: transaction throughput, transaction confirmation time, transaction fees, etc. 
    \item \textbf{Smart Contract Vulnerability Detection}. Given the immutability property of blockchains, vulnerability in smart contracts and DApps can have significant and dangerous consequences. Hence, studies have addressed the evaluation and detection of vulnerability in smart contracts. 
\end{itemize}

Table \ref{tab:distribution-UC} shows a summary of the distribution of the selected studies by use case.

\begin{table}[htbp]
\caption{Distribution of Studies by Use Case.}\label{tab:distribution-UC}
\begin{tabularx}{\textwidth}{l*{12}{>{\raggedright\arraybackslash}X}}
\toprule
\textbf{Use Case} & \textbf{2013} & \textbf{2014} & \textbf{2015} & \textbf{2016} & \textbf{2017} & \textbf{2018} & \textbf{2019} & \textbf{2020} & \textbf{2021} & \textbf{2022} & \textbf{2023} & \textbf{Total} \\
\midrule
\textbf{Address Classification} & 1 & 0 & 1 & 0 & 2 & 9 & 4 & 3 & 4 & 4 & 0 & 28\\
\textbf{Anomaly Detection} & 0 & 0 & 1 & 2 & 1 & 2 & 8 & 17 & 22 & 20 & 6 & 79\\
\textbf{Price Prediction} & 0 & 0 & 3 & 0 & 1 & 4 & 5 & 5 & 4 & 7 & 1 & 30\\
\textbf{Performance Prediction} & 0 & 0 & 0 & 0 & 0 & 0 & 1 & 1 & 3 & 2 & 0 & 7\\
\textbf{Vulnerability Detection} & 0 & 0 & 0 & 0 & 0 & 0 & 3 & 4 & 3 & 2 & 0 & 12\\
\textbf{Other} & 0 & 0 & 0 & 0 & 2 & 0 & 0 & 0 & 0 & 1 & 0 & 3\\
\hline
\textbf{Total} & 1 & 0 & 5 & 2 & 6 & 15 & 21 & 30 & 36 & 36 & 7 & 159\\
\bottomrule
\end{tabularx}
\end{table}

\subsection{Blockchain and Distribution}
The studies applied machine learning to data generated by various blockchains: 
\begin{itemize}
    \item \textbf{Bitcoin}. The blockchain data analyzed in the paper were coming solely from the Bitcoin blockchain. 
    \item \textbf{Ethereum}. The blockchain data analyzed in the paper were coming solely from the Ethereum blockchain. 
    \item \textbf{Bitcoin and Ethereum}. The blockchain data analyzed in the paper were coming from both the Bitcoin and Ethereum blockchain. 
    \item \textbf{Multiple}. The blockchain data analyzed in the paper were coming from multiple blockchains (other than the combination of Bitcoin and Ethereum). 
\end{itemize}

Table \ref{tab:distribution-blockchain} shows a summary of the distribution of the selected studies by blockchain. 

\begin{table}[htbp]
\caption{Distribution of Studies by Blockchain.}\label{tab:distribution-blockchain}
\begin{tabularx}{\textwidth}{l*{12}{>{\raggedright\arraybackslash}X}}
\toprule
\textbf{Blockchain} & \textbf{2013} & \textbf{2014} & \textbf{2015} & \textbf{2016} & \textbf{2017} & \textbf{2018} & \textbf{2019} & \textbf{2020} & \textbf{2021} & \textbf{2022} & \textbf{2023} & \textbf{Total} \\
\midrule
\textbf{Bitcoin} & 1 & 0 & 5 & 2 & 4 & 13 & 12 & 13 & 11 & 12 & 2 & 75\\
\textbf{Ethereum} & 0 & 0 & 0 & 0 & 1 & 1 & 9 & 16 & 17 & 19 & 5 & 68\\
\textbf{Bitcoin and Ethereum} & 0 & 0 & 0 & 0 & 0 & 0 & 0 & 1 & 3 & 2 & 0 & 6\\
\textbf{Multiple} & 0 & 0 & 0 & 0 & 0 & 0 & 0 & 0 & 2 & 1 & 0 & 3\\
\textbf{Other} & 0 & 0 & 0 & 0 & 1 & 1 & 0 & 0 & 3 & 2 & 0 & 7\\
\hline
\textbf{Total} & 1 & 0 & 5 & 2 & 6 & 15 & 21 & 30 & 36 & 36 & 7 & 159\\
\bottomrule
\end{tabularx}
\end{table}

\subsection{Data and Distribution}
The studies in this paper used different data sources: 
\begin{itemize}
    \item \textbf{Blockchain}. The authors extracted the data directly from the blockchain they studied, whether it is Bitcoin or Ethereum. 
    \item \textbf{Website}. The authors extracted the data from a blockchain explorer website, such as \url{blockchain.com}, \url{etherscan.io}, \url{walletexplorer.com}, or \url{xblock.pro}. 
    \item \textbf{Published Dataset}. The authors used a public dataset made available on a popular platform such as Kaggle, or made available through another type of data repository. 
    \item \textbf{Multiple}. The authors used various sources to construct a new dataset. 
\end{itemize}

The datasets also consisted of a wide range of data points. We created the following ranges of data points: 
\begin{itemize}
    \item \textbf{$<$ 1,000}. The dataset consists of less than 1,000 data points. 
    \item \textbf{1,000-2,000}. The dataset consists of more than 1,000 data points but less than 2,000 data points. 
    \item \textbf{2,000-5,000}. The dataset consists of more than 2,000 data points but less than 5,000 data points. 
    \item \textbf{5,000-10,000}. The dataset consists of more than 5,000 data points but less than 10,000 data points. 
    \item \textbf{10,000-50,000}. The dataset consists of more than 10,000 data points but less than 50,000 data points. 
    \item \textbf{50,000-100,000}. The dataset consists of more than 50,000 data points but less than 100,000 data points. 
    \item \textbf{100,000-500,000}. The dataset consists of more than 100,000 data points but less than 500,000 data points. 
    \item \textbf{500,000-1,000,000}. The dataset consists of more than 500,000 data points but less than 1,000,000 data points. 
    \item \textbf{$>$  1,000,000}. The dataset consists of more than 1,000,000 data points. 
\end{itemize}

Finally, we also recorded the availability of the dataset, i.e. whether or not the authors made the data used in their study available: 
\begin{itemize}
    \item \textbf{Yes}. The authors published or provided a link to the dataset they created and used for their studies.  
    \item \textbf{Yes (ext)}. The dataset was already public. Hence, the dataset is (already) available. 
    \item \textbf{Upon request}. The authors added a statement in their study regarding the availability of their dataset following a (sometimes reasonable) request. 
    \item \textbf{Partially}. A part of the dataset is available. It can happen when the authors take advantage of a public dataset and augment it with their own features. 
    \item \textbf{No}. The authors did not mention in their paper - whether in the core of the text, in a footnote or in a statement at the end of the article - that the dataset was available. When no such indication was mentioned, we considered that the data were not available. 
\end{itemize}

Tables \ref{tab:distribution-data-sources} to \ref{tab:distribution-availability} show a summary of the distribution of the selected studies by data. 

\begin{table}[htbp]
\caption{Distribution of Studies by Data Sources.}\label{tab:distribution-data-sources}
\begin{tabularx}{\textwidth}{l*{12}{>{\raggedright\arraybackslash}X}}
\toprule
\textbf{Data Source} & \textbf{2013} & \textbf{2014} & \textbf{2015} & \textbf{2016} & \textbf{2017} & \textbf{2018} & \textbf{2019} & \textbf{2020} & \textbf{2021} & \textbf{2022} & \textbf{2023} & \textbf{Total} \\
\midrule
\textbf{Bitcoin} & 0 & 0 & 1 & 0 & 0 & 1 & 1 & 1 & 0 & 1 & 1 & 6\\
\textbf{Blockchain.com} & 0 & 0 & 2 & 0 & 2 & 3 & 2 & 1 & 2 & 1 & 0 & 13\\
\textbf{Ethereum} & 0 & 0 & 0 & 0 & 0 & 0 & 2 & 2 & 0 & 0 & 0 & 4\\
\textbf{Etherscan.io} & 0 & 0 & 0 & 0 & 1 & 1 & 3 & 5 & 5 & 6 & 2 & 23\\
\textbf{Walletexplorer.com} & 0 & 0 & 0 & 0 & 0 & 1 & 2 & 1 & 1 & 0 & 0 & 5\\
\textbf{XBlock.pro} & 0 & 0 & 0 & 0 & 0 & 0 & 0 & 1 & 0 & 1 & 0 & 2\\
\textbf{Public Dataset} & 0 & 0 & 1 & 1 & 0 & 1 & 3 & 5 & 9 & 7 & 0 & 27\\
\textbf{Multiple} & 1 & 0 & 0 & 0 & 1 & 5 & 5 & 9 & 12 & 11 & 3 & 47\\
\textbf{Other} & 0 & 0 & 1 & 0 & 0 & 3 & 2 & 4 & 3 & 6 & 1 & 20\\
\textbf{Unclear} & 0 & 0 & 0 & 1 & 2 & 0 & 1 & 1 & 4 & 3 & 0 & 12\\
\hline
\textbf{Total} & 1 & 0 & 5 & 2 & 6 & 15 & 21 & 30 & 36 & 36 & 7 & 159 \\
\bottomrule
\end{tabularx}
\end{table}

\begin{table}[htbp]
\caption{Distribution of Studies by Number of Data Points.}\label{tab:distribution-data-points}
\begin{tabularx}{\textwidth}{l*{12}{>{\raggedright\arraybackslash}X}}
\toprule
\textbf{Data Points} & \textbf{2013} & \textbf{2014} & \textbf{2015} & \textbf{2016} & \textbf{2017} & \textbf{2018} & \textbf{2019} & \textbf{2020} & \textbf{2021} & \textbf{2022} & \textbf{2023} & \textbf{Total} \\
\midrule
$<$ 1,000 & 0 & 0 & 0 & 0 & 1 & 0 & 0 & 0 & 1 & 1 & 0 & 3\\
1,000-2,000 & 0 & 0 & 0 & 0 & 2 & 0 & 0 & 1 & 1 & 2 & 0 & 6\\
2,000-5,000 & 0 & 0 & 0 & 0 & 0 & 0 & 1 & 3 & 4 & 1 & 0 & 9\\
5,000-10,000 & 0 & 0 & 0 & 0 & 0 & 1 & 2 & 0 & 3 & 0 & 3 & 9\\
10,000-50,000 & 0 & 0 & 0 & 0 & 0 & 0 & 3 & 4 & 3 & 5 & 1 & 16\\
50,000-100,000 & 0 & 0 & 0 & 0 & 1 & 0 & 1 & 0 & 3 & 3 & 0 & 8\\
100,000-500,000 & 0 & 0 & 0 & 0 & 0 & 3 & 1 & 3 & 7 & 5 & 1 & 20\\
500,000-1,000,000 & 0 & 0 & 0 & 0 & 0 & 0 & 0 & 3 & 0 & 1 & 0 & 4\\
$>$ 1,000,000 & 1 & 0 & 4 & 2 & 0 & 6 & 7 & 11 & 9 & 9 & 1 & 50\\
Unclear & 0 & 0 & 1 & 0 & 2 & 5 & 6 & 5 & 5 & 9 & 1 & 34\\
\hline
Total & 1 & 0 & 5 & 2 & 6 & 15 & 21 & 30 & 36 & 36 & 7 & 159\\
\bottomrule
\end{tabularx}
\end{table}

\begin{table}[htbp]
\caption{Distribution of Studies by Data Availability.}\label{tab:distribution-availability}
\begin{tabularx}{\textwidth}{l*{12}{>{\raggedright\arraybackslash}X}}
\toprule
\textbf{Data Availability} & \textbf{2013} & \textbf{2014} & \textbf{2015} & \textbf{2016} & \textbf{2017} & \textbf{2018} & \textbf{2019} & \textbf{2020} & \textbf{2021} & \textbf{2022} & \textbf{2023} & \textbf{Total} \\
\midrule
Yes & 0 & 0 & 0 & 0 & 1 & 5 & 3 & 2 & 7 & 8 & 3 & 29\\
Yes (ext) & 0 & 0 & 1 & 1 & 0 & 1 & 3 & 8 & 12 & 10 & 0 & 36\\
Upon request & 0 & 0 & 0 & 0 & 0 & 0 & 0 & 0 & 1 & 2 & 3 & 6\\
Partially & 0 & 0 & 0 & 0 & 0 & 0 & 0 & 0 & 0 & 1 & 0 & 1\\
No & 1 & 0 & 4 & 1 & 5 & 9 & 15 & 20 & 16 & 15 & 1 & 87\\
\hline
Total & 1 & 0 & 5 & 2 & 6 & 15 & 21 & 30 & 36 & 36 & 7 & 159\\
\bottomrule
\end{tabularx}
\end{table}

\subsection{Machine Learning and Distribution}
The studies applied different types of analysis to the blockchain data: 
\begin{itemize}
    \item \textbf{Classification}. The paper addressed the application of a supervised learning approach and attempted to predict a label. 
    \item \textbf{Clustering}. The paper addressed the application of an unsupervised learning approach and attempted to group data points together. 
    \item \textbf{Deep Learning}.  The paper addressed the application of a neural network with multiple layers. 
    \item \textbf{Regression}. The paper addressed the application of a supervised learning approach and attempted to predict a continuous value.
    \item \textbf{Time Series Analysis}. The paper aimed to analyze a time series.
    \item \textbf{Combination}. The paper addressed the application of at least two algorithms belonging to the aforementioned categories. 
\end{itemize}

Table \ref{tab:distribution-ml} shows a summary of the distribution of the selected studies by machine learning type. 

\begin{table}[htbp]
\caption{Distribution of Studies by Model.}\label{tab:distribution-ml}
\begin{tabularx}{\textwidth}{l*{12}{>{\raggedright\arraybackslash}X}}
\toprule
\textbf{Model} & \textbf{2013} & \textbf{2014} & \textbf{2015} & \textbf{2016} & \textbf{2017} & \textbf{2018} & \textbf{2019} & \textbf{2020} & \textbf{2021} & \textbf{2022} & \textbf{2023} & \textbf{Total} \\
\midrule
Classification & 0 & 0 & 1 & 0 & 3 & 7 & 11 & 14 & 24 & 13 & 1 & 74\\
Clustering & 1 & 0 & 1 & 1 & 2 & 2 & 1 & 2 & 1 & 2 & 1 & 14\\
Deep Learning & 0 & 0 & 0 & 0 & 0 & 0 & 2 & 5 & 3 & 2 & 0 & 12\\
Regression & 0 & 0 & 0 & 0 & 0 & 3 & 0 & 1 & 1 & 3 & 0 & 8\\
Time Series Analysis & 0 & 0 & 3 & 0 & 1 & 0 & 2 & 0 & 0 & 1 & 0 & 7\\
Combination & 0 & 0 & 0 & 1 & 0 & 3 & 3 & 8 & 7 & 14 & 4 & 40\\
Other & 0 & 0 & 0 & 0 & 0 & 0 & 2 & 0 & 0 & 1 & 1 & 4\\
\hline
Total & 1 & 0 & 5 & 2 & 6 & 15 & 21 & 30 & 36 & 36 & 7 & 159\\
\bottomrule
\end{tabularx}
\end{table}

\section{Mapping Results} \label{sec:mapping-results}
\begin{figure}
	\centering
		\includegraphics[scale=0.6]{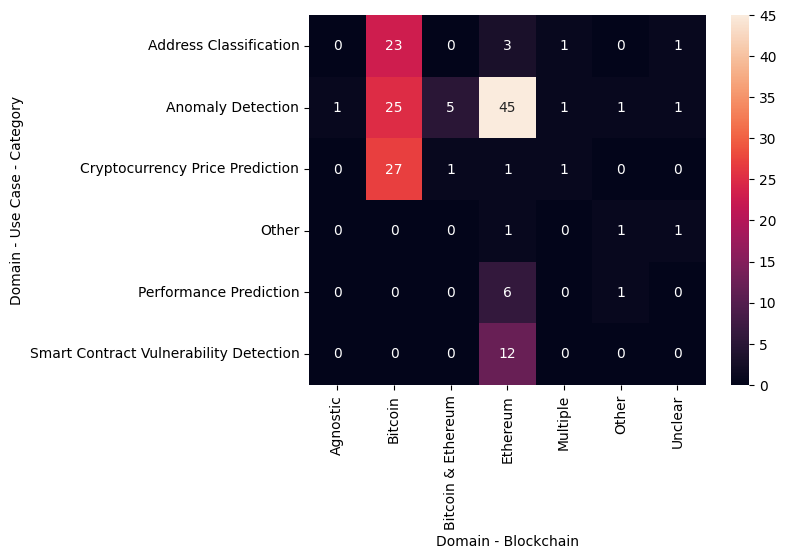}
	\caption{Distribution papers by Use Case and by Blockchain}
	\label{fig:heatmap-domai-blockchain}
\end{figure}

\begin{figure}
	\centering
		\includegraphics[scale=0.6]{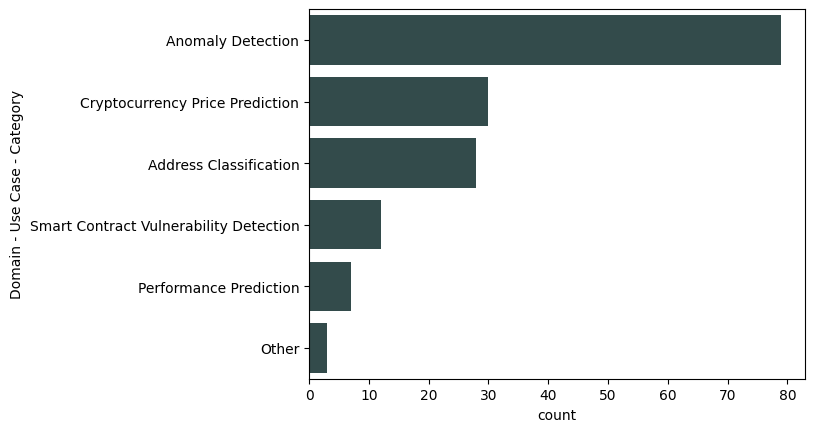}
	\caption{Distribution of papers by Use Case}
	\label{fig:Distribution of papers by Use Case}
\end{figure}

\begin{figure}
	\centering
		\includegraphics[scale=0.6]{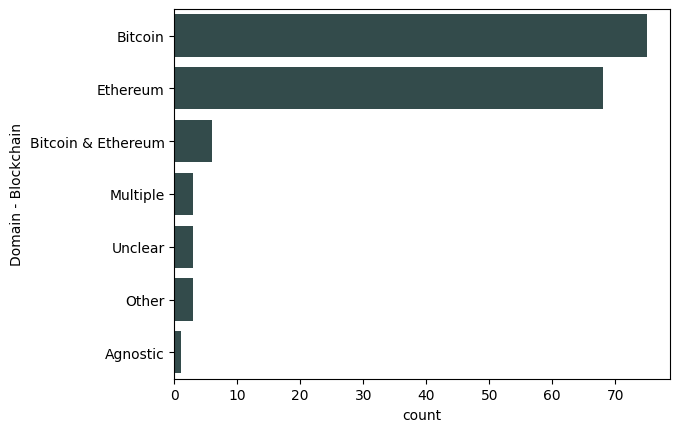}
	\caption{Distribution of papers by Blockchain}
	\label{fig:Distribution of papers by Blockchain}
\end{figure}

\begin{figure}
	\centering
		\includegraphics[scale=0.6]{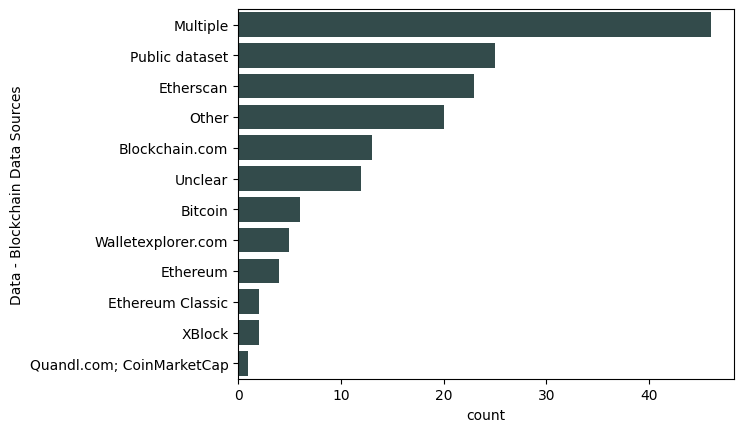}
	\caption{Blockchain Data Sources}
	\label{fig:Blockchain Data Sources}
\end{figure}

\begin{figure}
	\centering
		\includegraphics[scale=0.6]{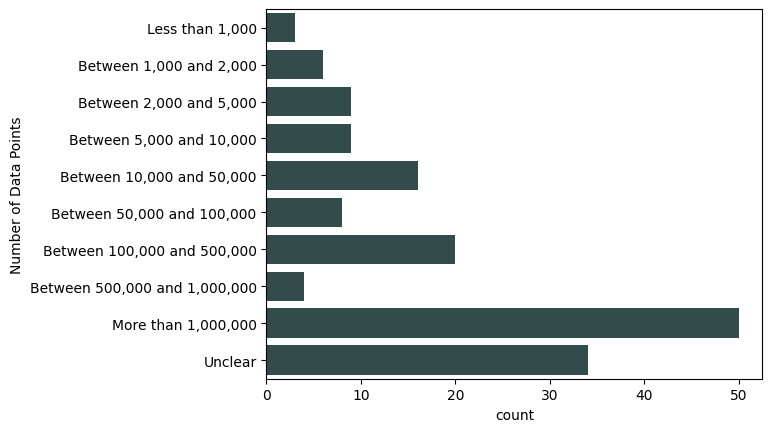}
	\caption{Blockchain Data Points Analyzed by Different Studies}
	\label{fig:Blockchain Data Points}
\end{figure}

\begin{figure}
	\centering
		\includegraphics[scale=0.6]{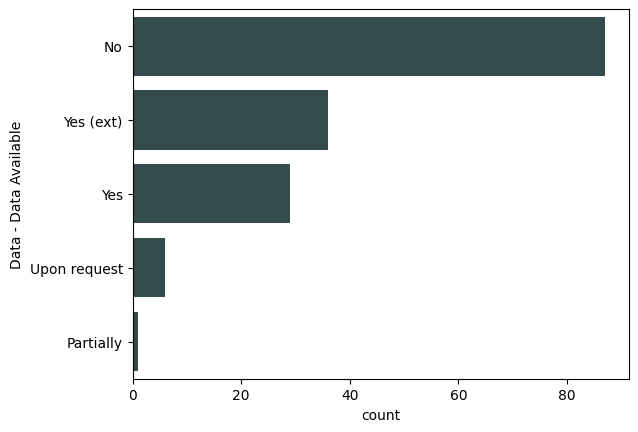}
	\caption{Distribution papers by Data Available}
    \label{fig:RQ2_Data_available}
\end{figure}

\begin{figure}
	\centering
		\includegraphics[scale=0.7]{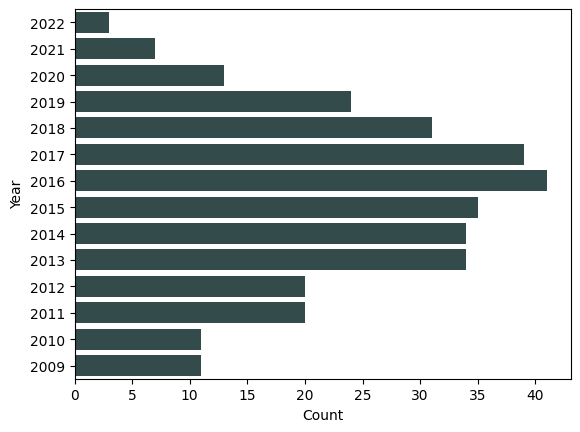}
	\caption{Distribution of papers by Period Covered}
    \label{fig:RQ2_Year_Included_in_Studies}
\end{figure}

\begin{figure}
        
	\centering
		\includegraphics[scale=0.6]{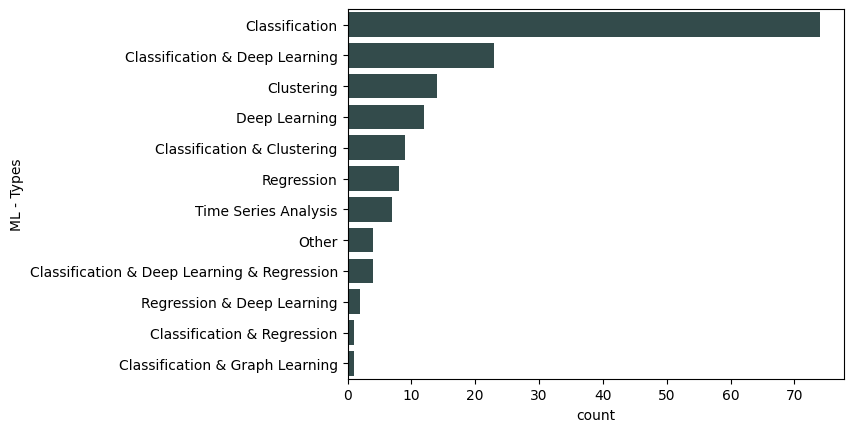}
	\caption{Distribution of papers by Type of Machine Learning Task}
    \label{fig:RQ3_ML_Task}
\end{figure}

\begin{figure}
	\centering
		\includegraphics[scale=0.6]{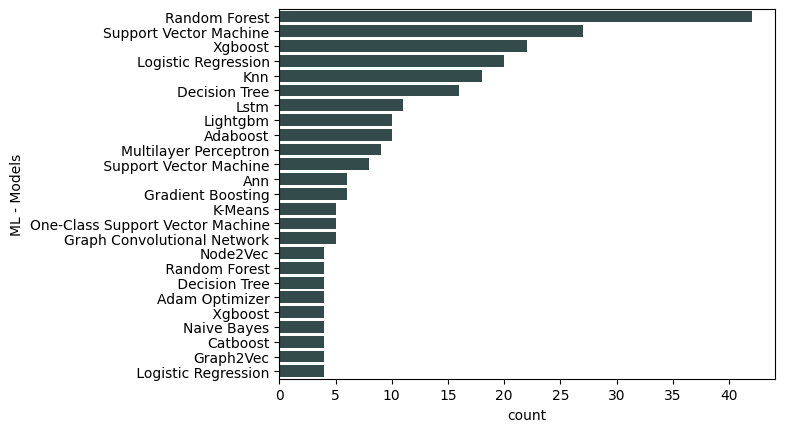}
	\caption{Distribution papers by Machine Learning Methods}
    \label{fig:RQ4_ML_Models}
\end{figure}

As mentioned before, the goals of this study are to provide an overview of the state of the art and to identify potential research gaps. We used heatmaps here that allow us to quickly grasp the dimensions explored by current research. The mapping results are represented in Figures \ref{fig:heatmap-domai-blockchain} to \ref{fig:RQ4_ML_Models}. 

The first dimension of our classification scheme, namely the Use Case, is discussed in more detail in Sections \ref{sec:address-classification} to \ref{sec:other}. The studies will be discussed by use case and we will address the other dimensions of our classification scheme as well, specifically the blockchain, the data, and the machine learning model. We can already state that, as shown in Figure \ref{fig:Distribution of papers by Use Case}, the majority of the studies we considered here addressed the problems of ``Anomaly Detection" (49.7\%), ``Cryptocurrency Price Prediction" (18.9\%), and ``Address Classification" (17.6\%). A smaller number of studies focused on ``Smart Contract Vulnerability Detection" (7.5\%) or ``Performance Prediction" (4.4\%). Three studies (1.9\%) were classified as ``Other". 
\cite{huang2017behavior} proposed a method for behavior pattern clustering for blockchain nodes; \cite{norvill2017automated} designed a solution to automatically label unknown contracts on Ethereum; and finally \cite{song2022eos} conducted some data analysis on EOS blockchain data. 

As far as the blockchains are concerned, we can see from Figure \ref{fig:Distribution of papers by Blockchain} that the majority of studies focused on a single particular blockchain, either Bitcoin (47.1\%) or Ethereum (42.8\%). Some authors worked on both (3.8\%) or on other multiple blockchains (1.9\%), e.g. Bitcoin, Ethereum and Ripple as in \cite{yae2022out} or EOSIO and Ethereum \cite{shen2021identity}. Finally, some studies analyzed the EOSIO blockchain \cite{song2022eos}, Hyperledger Fabric \cite{hang2022improved}, Steem \cite{kim2021posting}; or it was unclear which blockchain(s) was (were) analyzed. 

Next, Figure \ref{fig:Blockchain Data Sources} shows that the majority of the studies analyzed here used multiple data sources (29.6\%), such as Blockchain.com, Etherscan.io, a public dataset, and/or bitcoin clients. Other common sources include a single blockchain explorer website (27.0\%) such as Blockchain.com, Etherscan.io, Walletexplorer.com, or XBlock.pro; or a public dataset (17.0\%). 

We can also see from Figure \ref{fig:Blockchain Data Points} that most studies had a dataset consisting of more than 1,000,000 data points (31.4\%), between 100,000 and 500,000 data points (12.6\%), or between 10,000 and 50,000 data points (10.1\%). We should also note that a part of the studies we analyzed here (21.4\%) were unclear regarding the size of the dataset. 

Furthermore, Figure \ref{fig:RQ2_Data_available} displays the research data availability. Most studies did not share the data they used for the research (54.7\%), some were made available by the authors (18.2\%), and some were available by default because of their public characteristic (22.6\%). A small number of studies stated that the data were available on request (3.8\%), while the data for one study (0.6\%) were partially available.

Additionally, Figure \ref{fig:RQ2_Year_Included_in_Studies} shows that the period covered by the datasets fall mostly between 2013 and 2019. 

Finally, when we look at the models applied to the data, we can see in Figure  \ref{fig:RQ3_ML_Task} that a large majority of the studies conducted a classification (46.5\%). The second most popular processing was a combination of models (25.2\%), such as classification and clustering, classification and deep learning, or classification and regression for instance. Other frequent models include clustering alone (8.8\%) and regression alone (5.0\%). Some authors used time series analysis (4.4\%) or deep learning (7.5\%). 

As far as the specific algorithms are concerned, we can state from Figure \ref{fig:RQ4_ML_Models} that the Random Forest and the Support Vector Machine were the most popular ones in the studies we cover here, followed by the Logistic Regression, XGBoost, and K-Nearest Neighbors.

In the next subsections, we discuss each use case in more detail. For each use case, we provide a summary table reporting: the blockchain studied, the availability of the data and code (Data-Code), and the models and the performance measures used by the authors. We provide the best performance reported by the authors (in parenthesis) and the algorithms used to achieve that performance (in bold). Note that when no performance appears in the table, it means that either it was not explicitly mentioned, or that it was graphically represented in the paper. Also, when multiple ML models are highlighted, it means that some performed best on some metrics, while other algorithms performed best on some other measure. Finally, when experiments were carried out on various datasets, or when the performance measures were reported for the train and test sets; we decided to report only the best measure. When experiments were carried out on various blockchains, we provided the metrics for both platforms. 

\subsection{Address Classification - De-Anonymization} \label{sec:address-classification}

We found twenty-eight papers studying address classification. 

In our study, fourteen papers addressed de-anonymization or actors identification. Out of this set, fourteen papers focused on Bitcoin. 
Multiple authors applied a classification task using different algorithms: NB \cite{juhasz2018bayesian}; AdaBoost, Linear SVM, LR, Perceptron, RF \cite{ranshous2017exchange}; Multivariate Wald-Wolfowitz test \cite{monaco2015identifying}; AdaBoost, Bagging, DT, Extremely Randomized Trees, GB, NN, RF \cite{sun2019regulating}; AdaBoost, Bagging, DT, Extra Trees, GB, KNN, RF \cite{harlev2018breaking}. 
Various sets of features were fed to these algorithms: transaction-related features \cite{juhasz2018bayesian,monaco2015identifying,sun2019regulating,harlev2018breaking}; transaction-related features and graph-related features \cite{ranshous2017exchange}. 
The datasets used in these studies consisted of more than 1,000,000 data points \cite{juhasz2018bayesian,monaco2015identifying,sun2019regulating,harlev2018breaking}, and covered different periods: from September 29, 2011 to April 22, 2015 \cite{ranshous2017exchange}; and until April 7, 2013 \cite{monaco2015identifying}. 

Various authors addressed the problem of Bitcoin de-anonymization using a clustering algorithm and transaction-related features \cite{kang2020anonymization,remy2018tracking,meiklejohn2013fistful} . In \cite{zheng2018malicious,zheng2020identifying}, the authors improved the Louvain clustering algorithm in order to make Bitcoin de-anonymization possible, using a significant amount of Bitcoin transactions (almost 300 millions, and more than 120GB respectively) and two sets of features: transaction-related features and tracing data features. The other datasets consisted of more than 1,000,000 data points\cite{kang2020anonymization,remy2018tracking}, covering a period of about two months (March 10, 2020 to May 09, 2020) for \cite{kang2020anonymization}. 

Finally, four papers applied both Classification and Clustering: CatBoost, DT, XGboost \cite{tubino2022towards}; KNN \cite{shao2018identifying}; and RF and various clustering models (Contraction, SharedUser, K-Clique) \cite{zhang2018bitscope}. 
The datasets consisted of 23 features describing output transaction links \cite{tubino2022towards}; of address statistical features and address transaction history features \cite{shao2018identifying}; or on-chain data and off-chain data \cite{zhang2018bitscope}. Their size varied, from more than 1,000,000 data points \cite{zhang2018bitscope}, to less than 1,000,000 data points \cite{tubino2022towards}, and less than 500,000 \cite{shao2018identifying}; and covered large periods of time: January 3, 2009 to January 25, 2021 \cite{tubino2022towards}; and from January 2009 to September 2016 \cite{shao2018identifying}.
Also, \cite{xueshuo2021awap} proposed an Adaptive Weighted Attribute Propagation (AWAP) enhanced community detection model. For their solution, the authors selected 16 transaction-related features, and used a dataset consisting of less than 10,000 data points.  

\cite{shen2021identity} focused on identity inference on EOSIO and Ethereum, using a dataset consisting of less than 5,000 accounts. The authors applied GNN to subgraphs centered on target accounts.

In our study, eleven papers focused on address classification. 
The majority (seven papers) of them studied address classification on Bitcoin, using: ANN and RF \cite{lee2020machine}; Adaboost, DT, and imECOC \cite{liang2019targeted}; AdaBoost with DT, LightGBM, LR, MLP, NN, RF, SVM, XGBoost \cite{lin2019evaluation}; RF \cite{toyoda2018multi}; AdaBoost, GB, RF \cite{zola2019cascading}; ABC, BG, CART, ET, GB, KNN, LDA, LR, MLP, NB, RFC, SGD, SVM \cite{yin2017first}; and C4.5, ID3, KNN, PART \cite{blanco2022supervised}. 
As far as the selected features are concerned, \cite{lee2020machine} used transaction-related features, making the distinction between transmission and recipient addresses; \cite{liang2019targeted} used address-related features and network metrics; \cite{lin2019evaluation} used three types of features, namely basic statistics, extra statistics, and moments; \cite{toyoda2018multi,yin2017first,blanco2022supervised} extracted transaction history features; \cite{zola2019cascading}  used four sets of features, specifically entity features, address features, and two graph-related features. 
The periods covered by the datasets in the studies here: from January 1, 2016 \cite{lee2020machine}; from January 3rd, 2009 to June 30, 2018 \cite{lin2019evaluation}; from January 9, 2009 to February 9, 2017 \cite{toyoda2018multi}; from the Bitcoin genesis until February 2019 \cite{zola2019cascading}; and the period 2011-2018 \cite{blanco2022supervised}. These periods led to a dataset size of: less than 1,000,000 data points \cite{lee2020machine}; of more than 1,000,000 data points \cite{lin2019evaluation,zola2019cascading,blanco2022supervised,liang2019targeted}; of less than 500,000 data points \cite{toyoda2018multi}; less than 100,000 data points \cite{yin2017first}

One paper applied classification and deep learning to study the address classification, clustering and coin mixing problems on Bitcoin. Specifically, the authors in \cite{sun2022lstm} trained the following algorithms: LINE, LSTM Transaction Tree, RF using GCN Feature, SGD optimizer. The authors used sequence features extracted from transaction trees; and used a dataset of more than 1,000,000 data , covering the following periods: (i) November 16, 2013 to May 10, 2014; (ii) May 12, 2016 to May 17, 2016. 

Only two papers studied address classification on Ethereum. \cite{beres2021blockchain} focused on address clustering and de-anonymization, using a dataset of more than 1,000,000 data points, spanning from July 30, 2015 to April 4, 2020. The authors applied Node Embeddings. In \cite{liu2022fa}, the authors, using a dataset of more than 1,000,000 data points, applied a classification task and a deep learning task. They selected node features; and fed them to the following algorithms: Adam Optimizer, Cluster-GCN, Filter and Augment Graph Neural Network, GCN, GNN, GraphSAGE, H2GCN. 

Finally, \cite{tang2018learning} proposed PeerClassifier, a solution for blockchain peers classification, using the daily transaction amounts of peers extracted as a sequence. 

Other papers (two in our study) centered on agent characterization. \cite{jourdan2018characterizing} studied entities characterization on Bitcoin. The authors collected more than 1,000,000 data points, using blocks created before March 24 2018. The dataset was composed of five feature classes (address features, entity features, temporal features, centrality features, and motif features), and was fed to LightGBM and LR. \cite{liu2021characterizing} focused on characterizing key agents on Ethereum. The authors collected data between May 2, 2016 to June 29, 2019, for a total of less than 50,000 data points. They selected 28 features (categorized as volume features, temporal features, and structural features) and fed them to multiple classifiers (LightGBM, LR, MLP, RF, SVM). 

\begin{longtable}{p{3.5cm}p{1.5cm}p{1.5cm}p{4.5cm}p{3cm}}
\toprule
 Paper & Blockchain & Data-Code   & Model    & Performance\\ 
\midrule
\cite{tubino2022towards} 
    & Bitcoin 
    & N-N 
    & Classification: Decision Tree, \textbf{CatBoost}, \textbf{XGBoost}; and Clustering 
    & Clustering: aNMI (0.796), Completeness (0.6661), Homogeneity (1.0), Rand Index (0.532), V-score/NMI (0.796); Classification: ROC-AUC score (1.0) \\
\cite{xueshuo2021awap} 
    & Bitcoin 
    & Y-N 
    & Classification: \textbf{AWAP} (Enhanced Community Detection Model), BAGC, CP, DT, \textbf{LightGBM}, LR, RF, XGBoost; and Clustering: \textbf{AWAP} (Enhanced Community Detection Model), Big-CLAM, CESNA, Circles, CNM, CODICIL, CP, DeepWalk
    & Classification: Accuracy (0.95), F-Score (0.92), Precision (0.91); Clustering: F-Score (0.7583), Jaccard (0.5875), NMI (0.5683) \\
\cite{liu2021characterizing} 
    & Bitcoin 
    & Y-N 
    & Classification: LR, SVM, MLP, \textbf{RF}, LightGBM 
    & Accuracy (0.893), F1 (0.985), Precision (1.0), Recall (0.98), Macro F1 (0.865), Macro Precision (0.888), Macro Recall (0.862)\\
\cite{shen2021identity} 
    & EOSIO and Ethereum 
    & Y-N 
    & Classification and Deep Learning: \textbf{I$^2$BGNN} (Graph Neural Network), KNN, RF, SVM
    & F1 (0.9950), Precision (0.9917), Recall (0.9986)\\
\cite{jourdan2018characterizing} 
    & Bitcoin 
    & N-N 
    & Classification: \textbf{LightGBM}, LR
    & Accuracy (0.92), F1 (0.91), Precision (0.92) \\
\cite{juhasz2018bayesian} 
    & Bitcoin 
    & Y-N 
    & Classification: NB
    & Unclear \\
\cite{kang2020anonymization} 
    & Bitcoin
    & N-N  
    & Clustering: ``Multi-Input Addresses" heuristic
    & Reliability\\
\cite{lee2020machine} 
    & Bitcoin 
    & N-N
    & Classification: ANN, \textbf{RF} 
    & Accuracy (0.844) \\
\cite{liang2019targeted} 
    & Bitcoin 
    & N-N
    & Classification: Adaboost, DT-CART, \textbf{DT-HDDT}), \textbf{imECOC}
    & Accuracy (0.9147), AUC (0.9051), F1-measure (0.9117), G-mean (0.8283) \\
\cite{meiklejohn2013fistful} 
    & Bitcoin 
    & N-N
    & Clustering
    & Unclear \\
\cite{ranshous2017exchange} 
    & Bitcoin 
    & N-N
    & Classification: AdaBoost, Linear SVM, LR, Perceptron, \textbf{RF}
    & F1 (0.9973), Precision (0.9971), Recall (0.9976)\\
\cite{remy2018tracking} 
    & Bitcoin 
    & Y-N
    & Clustering
    & aNMI (0.67), F1 (0.86), NMI (0.89), Precision (0.98), Recall (0.91)\\
\cite{shao2018identifying} 
    & Unclear 
    & N-N
    & Classification: KNN; and Clustering
    & Accuracy (0.911), F1 (0.787), False Accept Rate (0.051), Precision (0.813), Recall (0.772), Validation rate (0,869)\\
\cite{tang2018learning} 
    & Unclear 
    & N-N
    & Classification: DT, KNN, \textbf{PeerClassifier}, SVM
    & Accuracy (0.745)\\
\cite{zhang2018bitscope} 
    & Bitcoin 
    & N-N
    & Classification: RF; and Clustering: Baseline, Contraction, \textbf{K-Clique}, SharedUser
    & Classification: Accuracy (0.738), F1 (0.701), Precision (0,769), Recall (0.654); Clustering: aMI (0.2394), Rand index adjusted for chance (0.03), V-measure (0.5229) \\
\cite{zheng2018malicious} 
    & Bitcoin 
    & N-N
    & Clustering
    & Accuracy (1.0), Comprehensiveness (1.0) \\
\cite{zheng2020identifying} 
    & Bitcoin
    & N-N
    & Clustering
    & Accuracy (0.91), Recall (1.0)\\
\cite{monaco2015identifying} 
    & Bitcoin 
    & Y-N
    & Classification: Multivariate Wald-Wolfowitz test
    & Accuracy (0.563), Equal Error Rate (0.084)\\
\cite{sun2019regulating} 
    & Bitcoin 
    & N-N
    & Classification: Adaptive Boosting, Bagging (Bootstrap Aggregation), DT, Extremely RT, \textbf{GB}, k-NN, RF
    & Accuracy (0.8042), F1 (0.7964), Precision (0.8055), Recall (0.8083), ROC Curve\\
\cite{harlev2018breaking} 
    & Bitcoin 
    & N-N
    & Classification: AdaBoost, Bagging Classifier, DT, Extra Trees, \textbf{GB}, KNN, RF
    & Accuracy (0.78), F1 (0.76), Precision (0.75), Recall (0.78), Support (451)\\
\cite{beres2021blockchain} 
    & Ethereum 
    & Y-Y
    & Classification: Node Embeddings
    & AUC, Average rank \\
\cite{lin2019evaluation} 
    & Bitcoin 
    & Y-N
    & Classification: AdaBoost, \textbf{LightGBM}, LR, \textbf{NN}, Perceptron, RF, SVM, XGBoost
    & Macro-F1 (0.86), Micro-F1  (0.91)\\
\cite{toyoda2018multi} 
    & Bitcoin 
    & Y-N
    & Classification: RF 
    & Accuracy (0.72)\\
\cite{zola2019cascading} 
    & Bitcoin 
    & N-N
    & Classification: Adaboost, \textbf{GB}, \textbf{RF}
    & Accuracy (0.9968), F1 (1.0), Matthews Correlation Coefficient (0.99), Precision (1.0), Recall (1.0), Standard Deviations\\
\cite{sun2022lstm} 
    & Bitcoin 
    & Y-Partially
    & Classification: Large-scale information network embedding , RF; and Deep Learning: \textbf{LSTM Transaction Tree}, \textbf{RF using GCN Feature}, SGD Optimizer
    & Classification: Accuracy, F1(0.983), Precision (0.982), Recall (0.990) \\
\cite{liu2022fa} 
    & Ethereum 
    & N-N
    & Classification and Deep Learning: Adam Optimizer, Cluster-GCN, \textbf{Filter and Augment GNN}, GCN, GraphSAGE, H2GCN
    & Macro-F1 (0.866), Macro-Precision (0.876), Macro-Recall (0.871), Micro-F1 (0.888) \\
\cite{yin2017first} 
    & Bitcoin 
    & N-N
    & Classification: AB, BG, CART, ET, \textbf{GB}, KNN, LDA, LR, MLP, NB, RF, SGD, SVM
    & Accuracy (0.8076), F1 (0.8056), Precision (0.8027), Recall (0.8271), Support (214) \\
\cite{blanco2022supervised} 
    & Bitcoin 
    & Y-N
    & Classification: C4.5, ID3 algorithm, KNN, PARTial Decision Tree (PART)
    & Accuracy, Kappa (Cohen’s Kappa)\\
\bottomrule
\caption{Papers Addressing Address Classification}
\end{longtable}
\label{tab:address-classification}

\subsection{Anomaly Detection} \label{sec:anomaly-detection}
In our study, we analyzed seventy-nine papers addressing Anomaly Detection: Ponzi Scheme Detection (10 papers), Phishing detection (12 papers), Ponzi Scheme and Phishing detection (1 paper), Intrusion or Attack Detection (4 papers), Illicit account detection (23 papers), Misbehavior or fraud detection (27 papers), and Bot detection (2 papers). 

In our study, we found ten papers working specifically on \textbf{Ponzi scheme detection}. All approaches focus on Ethereum, except for \cite{bartoletti2018data} who address the problem on Bitcoin. The authors built and made publicly available a dataset consisting of about 10,000 addresses and features of Bitcoin Ponzi schemes. 
As far as the papers focusing on Ethereum are concerned, authors used key account features and opcodes \cite{chen2018detecting,chen2019exploiting,wang2021ponzi,zhang2021detecting}, transaction features and code features \cite{jung2019data}, only the opcodes \cite{peng2020detection,fan2021spsd,aljofey2021supervised}, transaction features, features and bytecode \cite{zhang2021detecting}. 
The authors exploited a dataset consisting of: 1250 non-Ponzi scheme contracts and 131 Ponzi scheme contracts \cite{chen2018detecting,sun2020early}, 200 Ponzi scheme contracts and 3580 non-Ponzi scheme contracts \cite{chen2019exploiting,peng2020detection,aljofey2021supervised}, 3203 non-Ponzi scheme contracts and 172 Ponzi scheme contracts \cite{jung2019data}, 168 Ponzi schemes and 2851 normal smart contracts (the XBlock dataset) \cite{wang2021ponzi}, 386 Ponzi schemes and 3239 non-Ponzi schemes \cite{fan2021spsd}, 3614 Non-Ponzi contracts and 810 Ponzi contracts \cite{aljofey2021supervised}
The classifiers applied were: XGBoost\cite{chen2018detecting,peng2020detection,chen2019exploiting}, RF \cite{peng2020detection,aljofey2021supervised,jung2019data}, SGD and J48 \cite{jung2019data}, DT \cite{peng2020detection,chen2019exploiting,aljofey2021supervised}, Extremely Randomized Trees \cite{peng2020detection,aljofey2021supervised}, GB \cite{peng2020detection,aljofey2021supervised}, LightGBM \cite{peng2020detection,zhang2021detecting}, Logistic Regression \cite{peng2020detection}, SVM \cite{peng2020detection,chen2019exploiting}, Behavior Forest \cite{sun2020early}, Oversampling-based LSTM \cite{wang2021ponzi}, Ordered Boosting \cite{fan2021spsd}, AdaBoost, combination of Bagging-Tree and XGBoost, KNN \cite{aljofey2021supervised}, Isolation Forest \cite{chen2019exploiting}. 

Multiple authors addressed \textbf{phishing detection} on Ethereum, using classification and/or deep learning. They used different types of models: CT-GCN, Line graph and GCN, LR, One-Class SVM, SV, and XGBoost \cite{fu2022ct}; LightGBM, DeepWalk, Node2Vec, LINE, GCN \cite{chen2020phishing}; SVM \cite{yuan2020detecting}; Graph2Vec, Line Graph2Vec, Node2Vec, and WL-kernel \cite{yuan2020phishing}; IF, LR, NV, One-Class SVM \cite{wu2020phishers}; Graph2Vec, Node2Vec, Sub2Vec \cite{xia2022phishing}; AdaBoost, KNN, SVM \cite{wen2021transaction}; Adam optimizer, BPNN, DT, LightGBM, LR; LSTM, LSTM-FCN, Node2Vec, RF, RNN, SVM, Trans2Vec; \cite{wen2023novel}; GATNE-I (network embeddings), GNN, Random Walk\cite{wang2022heterogeneous}; DeepWalk, EGAT, GAT, GCN, Graph2Vec, GraphSageNode2Vec, MLP, Softmax (comparison with previous works where DElightGBM, LightGBM, MP-GCN, RF were used), Sub2Vec \cite{zhou2023detecting}; Adam optimizer, GCN, GIN, Graph2Vec, GraphSAGE, SF \cite{kim2023graph}; 48 Consolidated, C4.5 DT, Eth-PSD, Fast Decision Tree, JRip, NB Tree, OneR, PART Decision List \cite{kabla2022eth}. The inputs for these models were graph-based (transactional) features \cite{fu2022ct,chen2020phishing,yuan2020phishing,xia2022phishing,wang2022heterogeneous,zhou2023detecting,kim2023graph,wu2020phishers}, account features and network features \cite{wen2021transaction}, transaction features, state features, and transfer features \cite{wen2023novel}, and transaction-based features \cite{kabla2022eth,yuan2020detecting}. The datasets were composed of reconized phishing accounts and recognized non-phishing accounts; and were balanced \cite{fu2022ct,yuan2020phishing,zhou2023detecting} or unbalanced \cite{chen2020phishing,xia2022phishing,wen2021transaction,wen2023novel,kabla2022eth}. 

\cite{chen2021misbehavior} addressed both Ponzi schemes and phishing scams detection on Ethereum using data mining techniques. To address the first type of misbehavior, the authors extracted seven account features and used the opcodes of the smart contracts. They fed these features to an XGBoost algorithm and used a dataset consisting of 10,000 transactions. As far as the phishing scam detection problem is concerned, the authors used 7,795,044 transactions and compared multiple classifiers (DT, Dual-sampling Ensemble SVM, Dual-sampling Ensemble DT, Dual-sampling Ensemble lightGBM, LightGBM, and SVM). 


Multiple papers in our study addressed \textbf{fraud detection} on Bitcoin \cite{chen2019detecting,monamo2016unsupervised,monamo2016multifaceted,toyoda2017identification,lorenz2020machine,wahrstatter2023improving}, on Ethereum \cite{aljofey2022feature}, and on Bitcoin and Ethereum \cite{ashfaq2022machine,hall2022efficient,elmougy2021anomaly}. Various algorithms were used: an improved apriori algorithm \cite{chen2019detecting}; k-means and a trimmed k-means \cite{monamo2016unsupervised}; Boosted LR, kd-trees, maximum-likelihood based LR, RF, Trimmed k-means \cite{monamo2016multifaceted}; RF and XGBoost \cite{toyoda2017identification}, LR, RF, XGBoost, and  Angle-based Outlier Detection, Cluster-based Outlier Factor, Isolation Forest, KNN, Local Outlier Factor, One-Class SVM, and PCA \cite{lorenz2020machine}; k-means \cite{wahrstatter2023improving}; AB, BT, DT, Ensemble models, ET, GB, LGBM, RF, XGB \cite{aljofey2022feature}; RF, XGBoost \cite{ashfaq2022machine} ; Adam optimizer, GCN, MLP decoder \cite{hall2022efficient}; and LR, RF, SVM \cite{elmougy2021anomaly}. The Elliptic dataset \cite{weber2019anti} was used by  \cite{lorenz2020machine}  and used and augmented by \cite{hall2022efficient}. The sets of features selected or extracted by the authors include: transaction features \cite{chen2019detecting,toyoda2017identification,lorenz2020machine,ashfaq2022machine}; currency features, network features, and average neighborhood features \cite{monamo2016unsupervised,monamo2016multifaceted}; Turnover-features, connectivity-features, activity-features, utxos-specific-features \cite{wahrstatter2023improving}; opcode n-grams, transaction data, source code characters \cite{aljofey2022feature}. Except for \cite{lorenz2020machine,ashfaq2022machine,toyoda2017identification}, the dataset consisted of more than 1,000,000 data points.

Next, 17 papers addressed \textbf{abnormality detection} on Bitcoin \cite{lee2020toward,han2022research}, on Ethereum \cite{min2022portrait,patel2022evangcn,patel2020graph,al2021labeled,liu2022blockchain,scicchitano2020deep,hu2021transaction,podgorelec2019machine,baek2019model}, and on Bitcoin and Ethereum \cite{alarab2022effect,feldman2021bitcoin,chen2021bitcoin,rwibasira2022adobsvm,zhang2020anomaly,zhao2015graph}. The authors addressed the problem using classification, deep learning and/or clustering: ANN and RF \cite{lee2020toward}; GCN \cite{han2022research}; Extra Trees, GB, LR, MLP, RF, and XGBoost \cite{alarab2022effect}; AdaBoost, CatBoost, DT, KNN, LightGBM, LR, MLP, RF, SVC, and XGBoost \cite{feldman2021bitcoin}; AdaBoost, KNN, LOF, Mahalanobis distance-based method, MLP, OCSVM, RF, and SVM \cite{chen2021bitcoin}; KNN, NB, and SVM \cite{rwibasira2022adobsvm}; BTCOut, CTOuliers, DBSCAN, k-medoids algorithm, OddBall, Tclust \cite{zhang2020anomaly}; Apriori, Self Organised Maps \cite{min2022portrait}; k-means and RF \cite{baek2019model}; GCN, GNN, IForest, OC-GAT, OC-GCN, OC-SAGE, OCSVM \cite{patel2022evangcn,patel2020graph}; DT, Kernel SVM, KNN, LR, NB, RF, and One-class SVM \cite{al2021labeled}; CNN, GAT, GCN, Heterogeneous Graph Transformer Networks, RCNN, SVM \cite{liu2022blockchain}; sequence to sequence recurrent autoencoder \cite{scicchitano2020deep}; LSTM \cite{hu2021transaction}; DT, Isolation Forest and RF \cite{podgorelec2019machine}; and K-Means \cite{deepa2021cost}. 
The Elliptic dataset \cite{weber2019anti} was used by \cite{han2022research,alarab2022effect,feldman2021bitcoin} and was augmented for the authors studying Ethereum in addition to Bitcoin. The Ethereum Classic dataset \cite{farrugia2020detection} (composed of 2179 fraudulent accounts and 2502 normal accounts) was used in \cite{al2021labeled,scicchitano2020deep}. The size of the datasets used in these studies varied from more than 1,000,000 data points \cite{lee2020toward,chen2021bitcoin,zhang2020anomaly,zhao2015graph,min2022portrait,patel2020graph,scicchitano2020deep}, and less than 500,000 data points \cite{rwibasira2022adobsvm,baek2019model,patel2022evangcn,al2021labeled,liu2022blockchain,hu2021transaction,podgorelec2019machine}. The features used as input for the models include: transaction-based features \cite{lee2020toward,rwibasira2022adobsvm,al2021labeled,hu2021transaction,podgorelec2019machine}; graph-based transaction features or transactions depicted as graph \cite{chen2021bitcoin,zhang2020anomaly,zhao2015graph,patel2022evangcn,patel2020graph}; addresses (Smart Contract addresses and user addresses) and transactions \cite{min2022portrait}; features extracted from wallet transaction and wallets data \cite{baek2019model};  account and code features \cite{liu2022blockchain}; blocks features and transaction features \cite{scicchitano2020deep}; and time series transaction data \cite{deepa2021cost}. 


In our study, 23 papers addressed the \textbf{illicit account or illicit node detection} problem on Bitcoin \cite{wu2021detecting,li2020identifying,kanemura2019identification,elbaghdadi2021svm,alarab2022graph,eloul2021improving,boughaci2020enhancing,ahmed2021anti,alarab2020competence,zheng2021recognize} and on Ethereum \cite{ostapowicz2020detecting,farrugia2020detection,sun2019ethereum,vassallo2021application,weber2019anti,kilicc2022fraud,yuan2023eth,poursafaei2020detecting,ibrahim2021illicit,aziz2022lgbm,zhou2022prediction,bella2022detecting,poursafaei2021sigtran}. 
For this class of problem, various algorithms were used: DT, IF, IS, LR, OCSVM, PU learning \cite{wu2021detecting}; ANN, LSTM, RF, SVM with RBF kernel, and XBGoost \cite{li2020identifying}; GB and RF \cite{kanemura2019identification}; LR, RF, and SVM \cite{elbaghdadi2021svm}; Evolve-GCN, GCN, GCN and MLP, LR, MLP, RF, Skip-GCN, Temporal GCN \cite{alarab2022graph}; GCN, Graph Enhanced RF with Feedback model, Graphlet Spectral Correlation Analysis, RF; and Regression \cite{eloul2021improving}; AdaBoost, BN, k-means, NB, and RF \cite{boughaci2020enhancing}; CatBoost, LGBA, RF, XGBoost \cite{ahmed2021anti}; GCN, Node Embeddings, MLP \cite{alarab2020competence}; Attention Mechanism-based GCN, GCN, MLP, MP-GAT, Node Embeddings, Skip-GCN \cite{zheng2021recognize}; RF, SVM, and XGBoost \cite{ostapowicz2020detecting}; XGBoost \cite{farrugia2020detection}; Birch algorithm \cite{sun2019ethereum}; ASXGB, CatBoost, LGB, RF, XGBoost \cite{vassallo2021application}; DT, GB, KNN, MLP, RF, SVM \cite{kilicc2022fraud}; ETH Tracking Tree Method, LightGBM, LSTM; RF, XGBoost \cite{yuan2023eth}; Ensemble methods, LR, RF, and SVM \cite{poursafaei2020detecting}; DT (J48), KNN, and RF \cite{ibrahim2021illicit}; AdaBoost, KNN, LGBM, LR, MLP, RF, SVM, and XGBoost \cite{aziz2022lgbm}; C5, CatBoost, CAT Tree, LGB, NN, SVM; and Clustering \cite{zhou2022prediction}; LR, SVM, and XGBoost \cite{bella2022detecting}. Multiple authors used the Elliptic dataset to train and test their models \cite{elbaghdadi2021svm,alarab2022graph,eloul2021improving,boughaci2020enhancing,ahmed2021anti,alarab2020competence,zheng2021recognize,vassallo2021application} or the Ethereum Classic dataset \cite{ahmed2021anti,aziz2022lgbm,zhou2022prediction}, or another public dataset \cite{chen2020understanding} consisting of 10 millions of transactions \cite{sun2019ethereum}; otherwise, the authors collected their own data: \cite{wu2021detecting} collected 3 snapshots of 1,500,000 Bitcoin transactions; \cite{li2020identifying} used 24,720 illicit addresses and 1,209,850 licit addresses; \cite{kanemura2019identification} exploited 21 Darknet Market-related and 351 non-Darknet Market-related Bitcoin entities with the corresponding addresses and transactions; \cite{ostapowicz2020detecting} collected  2,200 wallets documented as involved in illegal activity and transactions from 349,999 randomly selected non-fraudulent wallets; \cite{kilicc2022fraud} collected blocks between 9,000,000 and 10,999,999, and 16,108,509 addresses and 141,242,377 transactions; \cite{yuan2023eth} used 5585 labeled addresses; \cite{poursafaei2020detecting} took advantage of 18,000,000 transactions; \cite{ibrahim2021illicit} used 7,809 transactions (7,651 normal and 159 fraudulent transactions); and \cite{bella2022detecting} used 2,600 illicit labeled addresses and 20,000 random addresses. 
As far as the features are concerned, the authors extracted and used: topological and temporal features \cite{li2020identifying}; transaction-based, address-based, block-based and statistics \cite{kanemura2019identification}; local features, local and aggregated features, local and node embeddings, and aggregated and node embeddings \cite{elbaghdadi2021svm,vassallo2021application}; local features \cite{alarab2022graph}; local and aggregated features \cite{alarab2020competence};  normal transaction-based features and erc20 token transaction-based features \cite{farrugia2020detection};  Ether trading transactions, smart contract creation and smart contract invocation and account-based features (external owned account and smart contract account) \cite{sun2019ethereum};  local and global features \cite{kilicc2022fraud}; transaction-based features and sequence features (information of the ETH flow) \cite{yuan2023eth}; general features, neighborhood features, local features, timestamp-related features \cite{poursafaei2020detecting}; transaction-based features \cite{ibrahim2021illicit,ostapowicz2020detecting}; account data, transactional data, and history data \cite{bella2022detecting}. 
Finally, \cite{poursafaei2021sigtran} proposed SigTran, a graph-based method for the detection of illicit nodes on Ethereum and Bitcoin. The authors used the dataset made public by \cite{weber2019anti} and extracted four sets of features: structural features, transactional features, regional features, and neighborhood features. They reported a better performance for their tool, compared to existing works. Specifically, the results for the Bitcoin and Ethereum blockchains respectively were: precision scores were 0.905 and 0.944, the recall scores were 0.947 and 0.940, the F1 scores were 0.918 and 0.942, the accuracy scores were 0.915 and 0.942, and the AUC scores were 0.976 and 0.976. 

Four papers in our study focus on \textbf{attack or intrusion detection} on Ethereum \cite{saveetha2022design,sanda2023long} and on Bitcoin, Bytecoin and Monero \cite{caprolu2021cryptomining}, using Classification and/or Deep Learning: LSTM and RNN \cite{saveetha2022design}; CNN, DT, KNN, and MLP \cite{sanda2023long}, RF \cite{caprolu2021cryptomining}; and DT, Ensemble methods, KNN, MLP, SGD, and SVM \cite{boudko2021predictive}. The datasets in these studies consisted of block-related features, gas-related features, and transaction-based features (Ethereum Classic dataset) \cite{saveetha2022design}; of transaction-based features, block-based features, blockchain-based features \cite{sanda2023long}; and of traffic-related features \cite{caprolu2021cryptomining}. 

Finally, a couple of papers worked on \textbf{bot detection}. 
\cite{zarpelao2019detection} proposed a solution for bot detection on the Bitcoin blockchain, using a One-class SVM . With a dataset consisting of more than 100,000 data points, and consisting of about 15 transaction-related features. \cite{kim2021posting}, on the other hand, worked on the same problem but focused on the Steem blockchain, and more specifically on posting bot detection on the Steem blockchain. The authors collected data about 984 accounts, divided into 325 bot accounts and 659 human accounts. They extracted various post-related features using the Minimum Average Cluster from Clustering Distance between Frequent words and Articles. Finally, they tested various classifiers (AdaBoost, DT, LightGBM, LSV, MLP, RF with Entropy, RF with Gini, XGBoost). 
\begin{longtable}{p{3.5cm}p{1.5cm}p{1.5cm}p{4.5cm}p{3cm}}
\toprule
 Paper & Blockchain & Data-Code   & Model    & Performance\\ 
\\ 
\midrule
\cite{saveetha2022design} 
    & Ethereum
    & N-N
    & Deep Learning: LSTM, RNN
    & Confusion matrix \\
\cite{rwibasira2022adobsvm} 
    & Bitcoin
    & N-N
    & Classification: KNN, NB, \textbf{SVM}
    & Energy value, Accuracy (0.98)\\
\cite{liu2022blockchain} 
    & Ethereum
    & Y-N
    & Classification and Deep Learning: CNN, GAT, GCN, \textbf{Heterogeneous Graph Transformer Networks}, RCNN, SVM
    & Macro-F1 (0.82), Micro-F1 (0.83)\\
\cite{wang2021ponzi} 
    & Ethereum
    & Y-N
    & Classification: Recurrent Neural Network - LSTM
    & F (0.96), Precision (0.97), Recall (0.96) \\
\cite{alarab2022effect} 
    & Bitcoin and Ethereum
    & Y-N
    & Classification: Extra Trees, GB, LR, MLP, \textbf{RF}, \textbf{XGBoost}
    & Accuracy (0.9802 - 0.9891), AUC (0.951 - 1.0), F1 (0.8239 - 0.9760)\\
\cite{caprolu2021cryptomining} 
    & Bitcoin, Bytecoin, Monero
    & N-N
    & Classification: \textbf{RF}
    & AUC (0.9908), F1 (0.96), FPR, TPR \\
\cite{fan2021spsd} 
    & Ethereum
    & N-N
    & Classification: \textbf{Ordered Boosting}
    & F (0.96), Precision (0.95), Recall (0.96) \\
\cite{farrugia2020detection} 
    & Ethereum
    & N-N
    & Classification: \textbf{XGBoost}
    & Accuracy (0.963), AUC (0.994) \\
\cite{aljofey2021supervised} 
    & Ethereum
    & N-N
    & Classification: AdaBoost, \textbf{combination of Bagging-Tree and XGBoost}, DT, Extra trees, GB, KNN, RF
    & Accuracy, AUC, F1 (0.95), FNR, FPR, Precision (0.97), Sensitivity (0.93), TNR, TPR \\
\cite{boudko2021predictive} 
    & Unclear
    & N-N
    & Classification: Ensemble methods, DT, KNN, MLP, SGD, SVM
    & Accuracy, Precision, Recall\\
\cite{chen2021misbehavior} 
    & Ethereum
    & Y-N
    & Classification: DT, Dual-sampling Ensemble DT, Dual-sampling Ensemble, \textbf{Dual-sampling Ensemble lightGBM}, LightGBM, SVM, XGBoost
    & AUC (- - 0.8097), F (0.86 - -), F1 (- - 0.8122), Precision (0.94 - 0.8196), Recall (0.81 - 0.8050)\\
\cite{deepa2021cost} 
    & Agnostic
    & N-N
    & Clustering: \textbf{K-Means}
    & Accuracy (0.982)\\
\cite{fu2022ct} 
    & Ethereum
    & Y-Y
    & Classification and Deep Learning: \textbf{CT-GCN}, Line graph and GCN, LR, One-Class SVM, SV, and XGBoost
    & Accuracy (0.8808), F1 (0.8814)\\
\cite{han2022research} 
    & Bitcoin
    & Y-N
    & Deep Learning: ARMAConv, \textbf{ClusterGCNConv}, GAT, GCN, GNN-FiLM, GraphConv, GraphSAGE, LEConv, Linear Regression, LR, MFConv, MLP, SGConv, SuperGATConv, \textbf{TAGCN}, TransformerConv
    & F1 (0.7298), Precision (0.9527), Recall (0.6334)\\
\cite{ostapowicz2020detecting} 
    & Ethereum
    & N-N
    & Classification: \textbf{RF}, \textbf{SVM}, \textbf{XGBoost}
    & F1 (0.4470), FN, FP, FPR (0.002), Precision (0.8571), Recall (0.8747), Specificity (0.9998), TN, TP \\
\cite{patel2020graph} 
    & Ethereum
    & Y-N
    & Deep Learning: IForest, \textbf{OC-GAT}, OC-GCN, \textbf{OC-SAGE}, OCSVM 
    & Accuracy (0.9075), AUROC, F1 (0.8346)\\
\cite{poursafaei2021sigtran} 
    & Bitcoin and Ethereum
    & Y-Y
    & Classification: Node2Vec, RiWalk, \textbf{SIGTRAN (LR)}
    & Accuracy (0.915 - 0.942), AUC (0.976 - 0.976), F1 (0.918 - 0.942), Precision (0.905 - 0.944), Recall (0.947 - 0.940)\\
\cite{scicchitano2020deep} 
    & Ethereum
    & Y-N
    & Deep Learning: Sequence to sequence recurrent autoencoder
    & Unclear\\
\cite{chen2018detecting} 
    & Ethereum
    & Y-N
    & Classification: \textbf{XGBoost}
    & Precision (0.94), Recall (0.81), F-score (0.86) \\
\cite{chen2019detecting} 
    & Bitcoin
    & Y-N
    & Frequent Set Mining: A Priori
    & - \\
\cite{chen2020phishing} 
    & Ethereum
    & N-N
    & Classification and Deep Learning: DeepWalk, \textbf{GCN}, LightGBM, LINE, Node2Vec
    & AUC (0.5866), F1 (0.2636)), Precision (7294), Recall (0.1735))\\
\cite{hu2021transaction} 
    & Ethereum
    & N-N
    & Deep Learning: \textbf{LSTM}
    & F1 (0.77), Precision (0.88), Recall (0.70)\\
\cite{li2020identifying} 
    & Bitcoin
    & N-N
    & Classification and Deep Learning: ANN, LSTM, \textbf{RF}, SVM with RBF kernel, \textbf{XGBoost}
    & F1 (0.9069), Precision (0.9567), Recall (0.8805)\\
\cite{peng2020detection} 
    & Ethereum
    & Y-N
    & Classification: DT, \textbf{Extremely Randomized Trees}, GB, LightGBM, LR, RF, SVM, XGBoost
    & F1 (0.95), Precision (0.98), Recall (of 0.93) \\
\cite{podgorelec2019machine} 
    & Ethereum
    & N-N
    & Outlier Detection: DT, Isolation Forest, RF
    & - \\
\cite{sun2019ethereum} 
    & Ethereum
    & Y-N
    & Clustering: Birch algorithm
    & - \\
\cite{wu2021detecting} 
    & Bitcoin
    & N-N
    & Classification: DT, IF, \textbf{IS}, LR, OCSVM, \textbf{PU learning}
    & FPR (0.0334), G-Mean (0.9479), TPR (0.9388)\\
\cite{yuan2020phishing} 
    & Ethereum
    & N-N
    & Classification: Node2Vec, Graph2Vec, \textbf{Line Graph2Vec}, WL-kernel
    & F1 (0.73), Precision (0.69), Recall (0.77)\\
\cite{zarpelao2019detection} 
    & Bitcoin
    & N-N
    & Classification: One-class SVM
    & AUC (0.99), FPR (0.01), TPR (1.0)\\
\cite{zhang2020anomaly} 
    & Bitcoin
    & Y-N
    & Classification: \textbf{BTCOut}, CTOuliers, DBSCAN, k-medoids algorithm, OddBall, Tclust
    & F2 (0.473), Precision (0.433), Recall (0.602)\\
\cite{zhao2015graph} 
    & Bitcoin
    & N-N
    & Clustering: Graph-based method
    & - \\
\cite{lee2020toward} 
    & Bitcoin
    & N-N
    & Classification: ANN, \textbf{RF}
    & F1 (0.9868) \\
\cite{kanemura2019identification} 
    & Bitcoin
    & N-N
    & Classification: GB, RF
    & F1, Precision, Recall \\
\cite{monamo2016unsupervised} 
    & Bitcoin
    & Y-N
    & Clustering: k-means, Trimmed k-means
    & Number known anomalies that were detected successfully\\
\cite{min2022portrait} 
    & Ethereum
    & Partially-Y
    & Clustering: Apriori, Self Organised Maps
    & - \\
\cite{poursafaei2020detecting} 
    & Ethereum
    & N-N
    & Classification: \textbf{AdaBoost}, LR, \textbf{RF}, \textbf{Stacking}, SVM
    & Accuracy (0.998), F1 (0.998), Precision (0.997), Recall (1.0) \\
\cite{ibrahim2021illicit} 
    & Ethereum
    & Y-N
    & Classification: DT (j48), \textbf{KNN}, RF
    & Accuracy (0.9877), F (0.987), Precision (0.986), Recall (0.988)\\
\cite{ashfaq2022machine} 
    & Bitcoin and Ethereum
    & Y-N
    & Classification: \textbf{RF}, XGboost
    & AUC (0.92), Precision, Recall \\
\cite{yuan2020detecting} 
    & Ethereum
    & N-N
    & Classification: \textbf{SVM}
    & F (0.846), Precision (0.871), Recall (0.822)\\
\cite{wu2020phishers} 
    & Ethereum
    & N-N
    & Classification: IF, LR, NV, \textbf{One-Class SVM}
    & F (0.908), Precision (0.927), Recall (0.893) \\
\cite{chen2019exploiting} 
    & Ethereum
    & Y-N
    & Classification: DT, \textbf{IF}, One-class SVM, SVM, XGBoost
    & F (0.79), Precision (0.95), Recall (0.69) \\
\cite{jung2019data} 
    & Ethereum
    & Y-N
    & Classification: \textbf{J48}, RF, \textbf{SGD}
    & F  (0.97), Precision (0.99), Recall (0.97)\\
\cite{zhang2021detecting} 
    & Ethereum
    & Y-Partially
    & Classification: \textbf{Improved LightGBM}
    & Accuracy, AUC (0.992), F (0.967), Precision (0.967), Recall (0.967) \\
\cite{sun2020early} 
    & Ethereum
    & Y-N
    & Classification: \textbf{Behavior Forest}
    & F, Precision (0.956), Recall (0.93)\\
\cite{ahmed2021anti} 
    & Bitcoin
    & Y-N
    & Classification: \textbf{CatBoost}, LGBA, \textbf{RF}, \textbf{XGBoost}
    & Accuracy (0.98), F1 (0.98), Recollect (0.976), Sensitivity (0.988)\\
\cite{bartoletti2018data} 
    & Bitcoin
    & Y-Y
    & Classification: Bayes Network, \textbf{RF}, RIPPER
    & Accuracy (0.988), AUC (0.978), F (0.443), G-mean (0.978), Precision (0.287), Recall (0.969), Specificity (0.987)\\
\cite{baek2019model} 
    & Ethereum
    & N-N
    & Classification: \textbf{RF}; and Clustering: k-Means
    & F1 (0.96), Precision (0.96), Recall (0.96), Support (9632)\\
\cite{toyoda2017identification} 
    & Bitcoin
    & N-N
    & Classification: RF, XGboost
    & F1, FPR, TPR \\
\cite{monamo2016multifaceted} 
    & Bitcoin
    & N-N
    & Classification: \textbf{Boosted LR}, kd-trees, Maximum-likelihood based LR, \textbf{RF}; and Clustering: Trimmed k-means
    & Kappa (0.99), Precision (0.99), Recall (1.0), ROC (0.99), Sensitivity\\
\cite{alarab2020competence} 
    & Bitcoin
    & Y-N
    & Classification and Deep Learning: \textbf{GCN}, MLP, Node Embeddings
    & Accuracy (0.974), F1 (0.773), Precision (0.899), Recall (0.678)\\
\cite{lorenz2020machine} 
    & Bitcoin
    & Y-Partially
    & Classification: LR, \textbf{RF}, XGBoost; and Anomaly Detection: Angle-based Outlier Detection, Cluster-based Outlier Factor, Isolation Forest, KNN, Local Outlier Factor, One-Class SVM, PCA
    & F1 (0.83)\\
\cite{aziz2022lgbm} 
    & Ethereum
    & Y-N
    & Classification: AdaBoost, KNN, \textbf{LGB}, LR, MLP, RF, \textbf{SVM}, XBoost; and Clustering
    & Accuracy (0.9860), F1 (0.9486), Precision (0.9948)\\
\cite{zhou2022prediction} 
    & Ethereum
    & Y-N
    & Classification: C5, \textbf{CatBoost}, CAT Tree, LGB, NN, SVM; and Clustering
    & Accuracy (0.947), AUC (0.9846)\\
\cite{zheng2021recognize} 
    & Bitcoin
    & Y-N
    & Classification and Deep Learning: Attention Mechanism-based GCN, GCN, MLP, \textbf{MP-GAT}, Node Embeddings, Skip-GCN
    & Accuracy (0.973), F1 (0.767), Precision (0.868), Recall (0.688) \\
\cite{bella2022detecting} 
    & Ethereum
    & N-N
    & Classification: LR, SVM, \textbf{XGBoost}
    & Accuracy (0.99), F1 (0.94), Precision (0.90), Recall (1.0)\\
\cite{patel2022evangcn} 
    & Ethereum
    & N-N
    & Classification and Deep Learning: \textbf{EVANGCN}, GCN, \textbf{OCGCN}
    & F1 (0.8702), MicroAvg F1 (0.7788), Precision (0.7284), Recall (0.6409)\\
\cite{hall2022efficient} 
    & Ethereum
    & Y-Y
    & Classification and Deep Learning: Adam optimizer, \textbf{GCN}, MLP
    & Accuracy (0.9813), F1 (0.7735), Precision (0.8225), Recall (0.8387)\\
\cite{vassallo2021application} 
    & Bitcoin and Ethereum
    & Y-N
    & Classification: \textbf{ASXGB}, CatBoost, LGB, \textbf{RF}, \textbf{XGBoost}
    & Accuracy (0.981 - 0.989), F1 (0.940 - 0.983), Precision (0.988 - 0.985), Recall (0.931 - 0.981)\\
\cite{aljofey2022feature} 
    & Ethereum
    & Y-N
    & Classification: AB, BT, DT, \textbf{Ensemble models}, ETC, GB, LGBM, \textbf{RF}, XGB
    & Accuracy (0.8967), F1 (0.8874), FNR (0.0018), FPR (0.1785), Precision (0.9837), Sensitivity (0.8148), TNR (0.9981), TPR (0.8214) \\
\cite{xia2022phishing} 
    & Ethereum
    & N-N
    & Classification: \textbf{Ego-Graph Embeddings}, Graph2Vec, Node2Vec, Sub2Vec
    & F1 (0.8199), Precision (0.8132), Recall (0.8271) \\
\cite{elbaghdadi2021svm} 
    & Bitcoin
    & Y-N
    & Classification: LR, \textbf{RF}, SVM
    & Accuracy (0.98851), Precision (0.65901), Recall (0.44866)\\
\cite{wen2021transaction} 
    & Ethereum
    & N-N
    & Classification: \textbf{AdaBoost}, KNN, \textbf{SVM}
    & Attack Success Rate, AUC (0.9276), Average number of Modified Edges, F1 (0.94), Precision (0.96), Recall (1.00) \\
\cite{wen2023novel} 
    & Ethereum
    & N-N
    & Classification: DT, LR, RF, SVM; and Deep Learning: Adam optimizer, LightGBM LSTM, \textbf{LBPS} (STM-FCN and BP neural network), Node2vec, RNN, Trans2vec
    & Accuracy (0.9730), F1 (0.9786), Precision (0.9813), Recall (0.9759)\\
\cite{alarab2022graph} 
    & Bitcoin
    & Y-N
    & Classification and Deep Learning: Evolve-GCN, GCN, GCN and MLP, LR, MLP, RF, Skip-GCN, \textbf{Temporal GCN}
    & Accuracy (0.977), F1 (0.806), Precision (0.927), Recall (0.713)\\
\cite{wang2022heterogeneous} 
    & Ethereum
    & N-N
    & Classification: DT, GATNE-I (Network Embeddings), GNN, LR, NB, \textbf{One-Class SVM}, Random Walk 
    & F1 (0.957), PR-AUC (0.889), ROC-AUC (0.959)\\
\cite{al2021labeled} 
    & Ethereum
    & Y-N
    & Classification: \textbf{DT}, Kernel SVM, KNN, LR, \textbf{NB}, One-class SVM, \textbf{RF}
    & Accuracy (0.99), Detection rate (0.953), FPR (0.0005)\\
\cite{eloul2021improving} 
    & Bitcoin
    & Y-N
    & Classification and Deep Learning: GCN, \textbf{Graph Enhanced RF with Feedback model}, Graphlet Spectral Correlation Analysis, RF; and Regression
    & F1 (0.821)\\
\cite{wahrstatter2023improving} 
    & Bitcoin
    & Y-N
    & Clustering: k-Means
    & CoinJoin-related metrics\\
\cite{kilicc2022fraud} 
    & Ethereum
    & Y-N
    & Classification: DT, \textbf{GB}, KNN, MLP, \textbf{RF}
    & Accuracy (0.985), F1 (0.743), Precision (0.682), Recall (0.876)\\
\cite{yuan2023eth} 
    & Ethereum
    & Y-Y
    & Classification and Deep Learning: ETH Tracking Tree Method, LightGBM, LSTM, RF, \textbf{XGBoost}
    & Accuracy (0.95), F1 (0.9542), Precision (0.961), Recall (0.9469) \\
\cite{zhou2023detecting} 
    & Ethereum
    & Y-N
    & Classification: DeepWalk, \textbf{EGAT}, GAT, GCN, Graph2vec, GraphSage, MLP, Node2vec, Softmax, Sub2vec
    & Accuracy (0.981), F1 (0.979), Precision (0.966), Recall (0.993) \\
\cite{kim2023graph} 
    & Ethereum
    & Y-N
    & Classification and Graph Learning: Adam Optimizer, GCN, Graph Isomorphism Network, Graph2Vec, \textbf{GraphSAGE}, SF
    & Accuracy (0.9878), AUROC (0.9876), F1 (0.9878), Precision (0.9878), Recall (0.9880)\\
\cite{sanda2023long} 
    & Ethereum
    & Y-N
    & Classification: DT, KNN, MLP; and Deep Learning: \textbf{CNN}
    & Accuracy (0.8754), F1 (0.85), Precision (0.8768), Recall (0.8448)\\
\cite{elmougy2021anomaly} 
    & Bitcoin and Ethereum
    & N-N
    & Classification: LR, \textbf{RF}, \textbf{SVM}
    & Accuracy (0.987-0.834), F1 (0.994-0.909), Recall (0.897-0.835) \\
\cite{kabla2022eth} 
    & Ethereum
    & N-N
    & Classification:  48 Consolidated, C4.5 DT, Eth-PSD, Fast DT, JRip, NB Tree, OneR, PART decision List
    & Accuracy (0.9776), F1 (0.97), FPR (0.01), Precision (0.97), Recall (0.97), ROC (0.97), Time taken to build a model (0.03)\\
\cite{kim2021posting} 
    & Steem
    & N-N
    & Classification: \textbf{AdaBoost}, DT, LightGBM, Linear Support Vector, MLP, \textbf{RF with Entropy}, \textbf{RF with Gini}, XGBoost
    & Accuracy (0.9268), F1 (0.8832), Precision (0.8512), Recall (0.9250)\\
\cite{boughaci2020enhancing} 
    & Bitcoin
    & Y-N
    & Classification: AdaBoost, Bayes Network, NB, \textbf{RF}; and Clustering: k-Means
    & FPR (0.007), Number of Correctly Classified Instances (0.9948), Number of Incorrectly Classified Instances (0.0052), PRC (1.0), Precision (0.995), Recall (0.995), ROC (1.0), TPR (0.995)  \\
\cite{feldman2021bitcoin} 
    & Bitcoin
    & Y-N
    & Classification: AdaBoost, CatBoost, DT, KNN, LightGBM, LGBM, LR, MLP, \textbf{RF}, SVC,  XGB, \textbf{XGBoost}
    & Accuracy (0.9921), F1 (0.957), Index of Balanced Accuracy (0.9599), Precision (0.997), Recall (0.922)\\
\cite{chen2021bitcoin} 
    & Bitcoin
    & Y-N
    & Classification: \textbf{AdaBoost}, KNN, LOF, Mahalanobis Distance-Based Method, MLP, OCSVM, \textbf{RF}, SVM
    & Accuracy (0.995), F1 (0.959), Precision (0.981), Recall (0.97)\\
\bottomrule
\caption{Papers Addressing Anomaly Detection}
\end{longtable}
\label{tab:anomaly-detection}

\subsection{Cryptocurrency Price Prediction} \label{sec:crypto-price-prediction}
We found thirty papers addressing cryptocurrency price prediction. 

The great majority of these studies focused on Bitcoin. 
Several authors applied a regression and/or deep learning, using different sets of algorithms: (i) ANN, LSTM, and RF \cite{chen2021machine}; Multiplayer Dynamic Game Model and SVM \cite{yan2022multi}; Bayesian NN, LR, and SVR \cite{jang2017empirical}; Graph Chainlets and RF \cite{akcora2018forecasting}; Bayesian Regression and a GLM/Random forest \cite{velankar2018bitcoin}; and a GRU, GRU-Dropout, GRU-Dropout-GRU, LSTM, and NN \cite{dutta2020gated}. 

The authors included the following features into their models: technology and economic factors \cite{chen2021machine,jang2017empirical,velankar2018bitcoin,dutta2020gated}, game theory-related factors \cite{yan2022multi}; and graph chainlets \cite{akcora2018forecasting}. 

The datasets used in these papers covered different periods of time: three different periods in \cite{chen2021machine} (specifically: (i) August 1, 2011 - December 31, 2013; (ii) August 1, 2013 - December 21, 2014; (iii) July 1, 2014 - December 31, 2017; (iv) July 1, 2015 - July 31, 2018); from September 13, 2011, to July 21, 2017 \cite{jang2017empirical}; from 2008 to 2009 \cite{akcora2018forecasting}; and from January 1, 2010 to June 30, 2019 \cite{dutta2020gated}

Six papers applied a time series analysis to Bitcoin data, with different periods of time: from October 2013 - March 2021 \cite{cai2022risk}; from January 2013–May 2017 \cite{poyser2019exploring}; from September 4, 2011 to February 28, 2014 \cite{kristoufek2015main}; from October 27, 2014 to January 12, 2015 \cite{georgoula2015using}; until December 31, 2014 \cite{yang2015bitcoin}. In \cite{li2017technology}, the authors studied two periods: (i)  January 1, 2011 to December 31, 2013; (ii) July 1, 2013 to December 31, 2014. 
The features used in these papers include both technology factors and economic factors. 

Nine papers addressed the price prediction problem from a classification and/or deep learning perspective, and tried to predict the direction of the price movement instead of the price point. 
The algorithms used in these papers are: Linear Discriminant Analysis, LR, LSTM, Quadratic Discriminant Analysis, RF, SVM, XGBoost \cite{chen2020bitcoin}; Ensemble, Feedforward NN, GB, GRU, LSTM, RNN, RF \cite{jaquart2021short}; a regression for classification task \cite{antulov2018inferring}; DT, KNN, Linear Discriminant Analysis, LR, Quadratic Discriminant Analysis, RF, SVM, XGBoost \cite{ranjan2022bitcoin}; CNB, CNN, CNN-LSTM, CNN-GRU, DT, GRU, LR, LSTM, SVM, RF \cite{kanji2022predicting}; BPNN, PCA-SVR, SDAE, and SVR \cite{liu2021forecasting}; CNN, Combinations of CNNs and RNNs, DNN, DRN, Ensemble Models, RNN and LSTM \cite{ji2019comparative};  DNDT, DRCNN, DSVR \cite{lamothe2020deep}; and CNN \cite{yogeshwaran2019project}. 
In order to train and test these models, the authors used various sets of features: technology features, economic features and attention features  \cite{chen2020bitcoin,jaquart2021short,ranjan2022bitcoin,kanji2022predicting,lamothe2020deep}; technology features and economic features \cite{liu2021forecasting,ji2019comparative,yogeshwaran2019project}; and Bitcoin daily transaction graphs \cite{antulov2018inferring}. 
Finally, the authors used data covering various periods of time: from July 2013 - December 2019 \cite{liu2021forecasting}; from November 29, 2011 to December 31, 2018 \cite{ji2019comparative}; from 2011 to 2019, with a quarterly frequency of data \cite{lamothe2020deep}; and from 2015 \cite{yogeshwaran2019project}. In \cite{jaquart2021short}, the authors used a dataset covering the period from March 4, 2019 to December 10, 2019. Two papers used two datasets: \cite{chen2020bitcoin} ((i) February 2 2017 to February 1 2019, and (ii) July 17 2017 to January 17 2018); and \cite{ranjan2022bitcoin} ((i) data and Bitcoin daily price from January 1, 2017, to December 31, 2019; and (ii) Bitcoin 5-min interval price for the period November 2, 2016 to June 11, 2018). The authors in \cite{kanji2022predicting} used a dataset of less than 100,000 data points. 

Four papers applied both a classification and a regression to address the cryptocurrency price prediction problem. The algorithms used in the papers we analyzed here include: ANN, Ensemble, RNN, and SVM \cite{mallqui2019predicting}; ARIMA, LSTM, and RNN \cite{mcnally2018predicting}; ANN, LSTM, SANN, and SVM \cite{mudassir2020time}; and Deep Cross Networks, DNN, and FLRDS \cite{nagula2022new}. 
The datasets used in the studies consisted of various sets of features: Bitcoin price-related features and transaction fees \cite{mallqui2019predicting}; and technology features and economic features \cite{mcnally2018predicting,mudassir2020time,nagula2022new}. 
The datasets used for these studies covered various numbers of periods: two periods ((i) August 19, 2013 - July 19, 2016; (ii) April 1st, 2013 - April 1st, 2017) in \cite{mallqui2019predicting}; one period (August 19, 2013 to July 19, 2016) in \cite{mcnally2018predicting} and (from February 2014 - September 2021) in \cite{nagula2022new}; and three periods ((i) April 1, 2013 to July 19, 2016; (ii) April 1, 2013 to April 1, 2017; (iii) April 1, 2013 to December 31, 2019) in \cite{mudassir2020time}. 

Finally, for the Bitcoin price prediction, \cite{chalkiadakis2022chain} used Multimodal Causality Testing with technology features, economic features and attention features, and data spanning from February 2018 - January 2020; while \cite{li2023bitcoin} used various algorithms (i.e. Multi-Window Prediction Framework, Full connected NN, and SVM). To get this result, the authors extracted transaction subgraphs and used data from April 1, 2013 to December 31, 2017. 

\cite{kim2021predicting} analyzed Ether price prediction, with data ranging from August 11, 2015 to November 28, 2018. They predicted the cryptocurrency price using SVM; ANN with technology features and economic features. 

One paper studied price prediction on Bitcoin and Ethereum \cite{saad2019toward}. The authors used technology features and economic features as input to several regressors (ANN, Conjugate Gradient, Elastic Net, GB, LASSO, LR, LSTM, RF). They used a dataset of less than 2,000 data points covering a period from December 18, 2017 - November 30, 2020. 

Finally, \cite{yae2022out} focused on cryptocurrency price prediction on Bitcoin, Ethereum, and Ripple. The authors used a dataset consisting of more than 1,000,000 data points and composed of the technology features, economic features and attention features. They used various regressors (Conjugate Gradient Approach, Linear Regression, LSTM, regression with GB, regression with RF).

\begin{longtable}{p{3.5cm}p{1.5cm}p{1.5cm}p{4.5cm}p{3cm}}
\toprule
 Paper & Blockchain & Data-Code   & Model    & Performance\\ 
\\ 
\midrule
\cite{kim2021predicting} 
    & Ethereum
    & N-N
    & Regression: \textbf{ANN}, SVM
    & MAPE (0.048), RMSE (0.068)\\
\cite{chalkiadakis2022chain} 
    & Bitcoin
    & Y-Y
    & Multimodal Causality Testing: Multiple-Output Convolutional Gaussian
    & Hypotheses Testing\\
\cite{cai2022risk} 
    & Bitcoin
    & N-N
    & Time Series Analysis: VAR
    & Unclear \\
\cite{chen2020bitcoin} 
    & Bitcoin
    & N-N
    & Classification and Deep Learning: Linear Discriminant Analysis, LR, \textbf{LSTM}, Quadratic Discriminant Analysis, RF, SVM, \textbf{XGBoost}
    & Accuracy (0.672), F1 (0.776), Precision (0.817), Recall (0.840)\\
\cite{chen2021machine} 
    & Bitcoin
    & N-N
    & Regression and Deep Learning: ANN, \textbf{LSTM}, RF
    & DA (76.5), MAE (8.7512), MAPE (2.2793), RMSE (12.0632) \\
\cite{jaquart2021short} 
    & Bitcoin
    & N-N
    & Classification and Deep Learning: Ensemble, Feedforward NN, GB, GRU, \textbf{LSTM}, RF, RNN
    & Accuracy (0.56)\\
\cite{mallqui2019predicting} 
    & Bitcoin
    & N-N
    & Classification: ANN, \textbf{Ensemble}, SVM; and Regression: ANN, RNN, \textbf{SVM}, 
    & Accuracy (0.6291), AUC (0.58), MAE (6.7), MAPE (1.14\%), RMSE (12.12)\\
\cite{liu2021forecasting} 
    & Bitcoin
    & N-N
    & Deep Learning: BPNN, PCA-SVR, \textbf{SDAE}, SVR
    & DA (0.5985), MAPE (0.1019), RMSE (160.63) \\
\cite{yae2022out} 
    & Bitcoin, Ethereum, and Ripple
    & N-N
    & Regression: ANN, Combination Forecasts, Elastic Net, GB, \textbf{LAD}, LASSO, Monitoring Forecasts, \textbf{Rank} Regression, RF, Shrinkage Estimators
    & Accuracy (0.561-0.538-0.557), CSSED, MAD (2.68-3.53-3.18), R$^2$ (0.0269-0.0171-0.0212)\\
\cite{yan2022multi} 
    & Bitcoin
    & N-N
    & Regression: \textbf{Multiplayer Dynamic Game Model}, \textbf{SVM}
    & Accuracy (0.642), Precision (0.877), Sensitivity (0.439), Specificity (0.623)\\
\cite{antulov2018inferring} 
    & Bitcoin
    & N-N
    & Classification: Regression for classification task
    & AUC PR (0.51), AUC ROC (0.73)
     \\
\cite{jang2017empirical} 
    & Bitcoin
    & N-N
    & Regression: \textbf{Bayesian NN}, Linear Regression, SVR
    & MAPE (0.0198-0.6302), RMSE (0.0244-0.5114) \\
\cite{poyser2019exploring} 
    & Bitcoin
    & N-N
    & Time Series Analysis: Bayesian Structural Time Series
    & MAE (12.139), MSE (457.65), sMAPE (2.970\%) \\
\cite{saad2019toward} 
    & Bitcoin and Ethereum
    & N-N
    & Regression: \textbf{Conjugate Gradient}, Linear Regression, LSTM, Regression with GB, Regression with RF
    & MAE (- - -), RMSE (- - -) \\
\cite{ji2019comparative} 
    & Bitcoin
    & N-N
    & Deep Learning: Combinations of CNNs and RNNs, CNN, \textbf{DNN}, Deep Residual Networks, Ensemble Models, Recurrent Neural Networks and LSTM, \textbf{SVM}
    & Accuracy (0.5306), F1 (0.67), MAPE (3.60\%), Precision (0.5290), Recall (1.0), Specificity (0.5370)\\
\cite{kristoufek2015main} 
    & Bitcoin
    & N-N
    & Time Series Analysis
    & Unclear \\
\cite{lamothe2020deep} 
    & Bitcoin
    & N-N
    & Deep Learning: DNDT, \textbf{DRCNN}, DSVR
    & Accuracy (0.9527), MAPE (0.29), RMSE (0.66) \\
\cite{li2017technology} 
    & Bitcoin
    & Y-N
    & Time Series Analysis: Time series using ARDL Model
    & F-Statistics, R$^2$ (0.284) \\
\cite{mcnally2018predicting} 
    & Bitcoin
    & N-N
    & Classification and Deep Learning and Regression: \textbf{ARIMA}, \textbf{LSTM}, \textbf{RNN}
    & Accuracy (0.5278), Precision (1.0), RMSE (5.45\%), Sensitivity (0.4040), Specificity (1.0)\\
\cite{mudassir2020time} 
    & Bitcoin
    & N-N
    & Classification and Deep Learning and Regression: ANN, \textbf{LSTM}, \textbf{SANN}, SVM
    & Accuracy (0.64), AUC (0.66), F1 (0.71), MAE (1.24), MAPE (0.52\%), RMSE (1.58) \\
\cite{yang2015bitcoin} 
    & Bitcoin
    & N-N
    & Time Series Analysis: Vector Autoregression
    & P-Value \\
\cite{akcora2018forecasting} 
    & Bitcoin
    & Y-Y
    & Regression: \textbf{Graph Chainlets}, RF
    & RMSE \\
\cite{velankar2018bitcoin} 
    & Bitcoin
    & N-N
    & Regression: Bayesian Regression, GLM/RF
    & Unclear\\
\cite{dutta2020gated} 
    & Bitcoin
    & N-N
    & Regression and Deep Learning: GRU, \textbf{GRU-Dropout}, GRU-Dropout-GRU, LSTM, Neural Network
    & RMSE (0.017)\\
\cite{yogeshwaran2019project} 
    & Bitcoin
    & N-N
    & Deep Learning: CNN
    & Unclear \\
\cite{georgoula2015using} 
    & Bitcoin
    & N-N
    & Time Series Analysis: OLS
    & p-Value \\
\cite{li2023bitcoin} 
    & Bitcoin
    & Y-N
    & Unclear: Full connected NN , \textbf{Multi-Window Prediction Framework}, SVM
    & MAPE (1.69\%)\\
\cite{nagula2022new} 
    & Bitcoin
    & N-N
    & Classification and Regression: DCN, DNN, \textbf{Hybrid Model}
    & Loss Ratio (0.0), MAE (9.96\% - 1707.42), ROI (11\%), Sharpe ratio (1.03), Total profit (USD 65,043.99), Volatility (0.07), Win Ratio (1.0), Win–loss Ratio (-) \\
\cite{ranjan2022bitcoin} 
    & Bitcoin
    & N-N
    & Classification: DT, KNN, \textbf{Linear Discriminant Analysis}, \textbf{LR}, Quadratic Discriminant Analysis, RF, \textbf{SVM}, XGBoost
    & Accuracy (0.648), Confusion Matrix, F1 (0.63), Precision (0.71), Recall (1.0)\\
\cite{kanji2022predicting} 
    & Bitcoin
    & Y-Y
    & Classification and Deep Learning: \textbf{CNN}, \textbf{CNN-GRU}, \textbf{CNN-LSTM}, Complement NB, DT, GRU, LR, LSTM, RF, SVM
    & Accuracy (0.6086), F1, Precision (0.60), Recall (0.60)\\
\bottomrule
\caption{Papers Addressing Cryptocurrency Price Prediction}
\end{longtable}
\label{tab:cryptocurrency-price-prediction}

\subsection{Performance Prediction}
Seven papers in this study addressed a performance prediction problem, using a classification task \cite{singh2019prediction,fajge2021wait,oliveira2021analyzing}, a regression task \cite{hang2022improved}, a classification and deep learning tasks \cite{lan2022gas,mars2021machine}, or a time series analysis \cite{pierro2019influence}. 

Two papers proposed a solution for gas price prediction on Ethereum. \cite{mars2021machine} used the gas price spanning from October 10, 2020 to October 24, 2020, while \cite{lan2022gas} used gas-related features spanning from March 17, 2022 to April 13, 2022. 

The five other papers focused on the transactions: transaction throughput prediction on Hyperledger Fabric \cite{hang2022improved}, transaction confirmation time prediction on Ethereum \cite{singh2019prediction,fajge2021wait,oliveira2021analyzing}, and transaction fees prediction \cite{pierro2019influence}. 
In order to address these topics, the authors used: transaction latency, send rate, transaction throughput, and error \cite{hang2022improved}; gas-related features and transaction-time related features \cite{singh2019prediction}; transaction-specific features, block- and network-specific features \cite{fajge2021wait}; and cryptocurrency prices-related features and block- and network-specific features \cite{pierro2019influence}. 

All datasets used in these papers consisted of at least more than 10,000 data points. They covered various periods, ranging from before November 2018 until April 2022. August 2019.

\begin{longtable}{p{3.5cm}p{1.5cm}p{1.5cm}p{3cm}p{4.5cm}}
\toprule
 Paper & Blockchain & Data-Code   & Model    & Performance\\ 
\\ 
\midrule
\cite{hang2022improved} 
    & Hyperledger Fabric
    & Y-N
    & Regression: ANN
    & MAD (0.250), MSE (6.701), RMSE (2.716) \\
\cite{singh2019prediction} 
    & Ethereum
    & N-N
    & Classification: \textbf{MLP}, NB, RF
    & Accuracy (0.8361), Cohen's Kappa (0.7877)\\
\cite{pierro2019influence} 
    & Ethereum
    & Y-N
    & Time Series Analysis: Granger causality
    & p-Value\\
\cite{fajge2021wait} 
    & Ethereum
    & N-N
    & Classification: Ensemble Approach, KNN, MLP, \textbf{RF}, SVM, 
    & Accuracy (0.9018), F1 (0.897), Macro F1, Precision (0.896), Recall (0.896), ROC (0.936)\\
\cite{lan2022gas} 
    & Ethereum
    & N-N
    & Classification and Deep Learning: \textbf{LSTM}, \textbf{XBoost}
    & MAE (0.064), RMSE (0.108), R$^2$ (0.962)\\
\cite{mars2021machine} 
    & Ethereum
    & Y-Y
    & Classification and Deep Learning: \textbf{GRU}, \textbf{LSTM}, Prophet, RNN
    & MAE (0.063), MSE (0.008), R$^2$ (0.896), RMSE (0.088)\\
\cite{oliveira2021analyzing} 
    & Ethereum
    & Y-Y
    & Classification: \textbf{DT}, LR, \textbf{RF}, SVM
    & Accuracy (0.98272), AUC-ROC (0.873668), F1 (0.674454), F2 (0.617612), Matthews Correlation Coefficient (0.702714), Precision (0.797683), Recall (0.973890), TNR (0.995), TPR (0.974)\\
\bottomrule
\caption{Papers Addressing Performance Prediction}
\end{longtable}
\label{tab:performance-prediction}

\subsection{Smart Contract Vulnerability Detection} \label{sec:smart-contract-vulnerability-detection}
In our studies, we found twelve papers addressing smart contract vulnerability detection, assessment or classification on Ethereum. All papers used classifiers and/or deep learning models and reported various performance measures. 

In the papers in our study, the authors experimented with a set of classifiers and/or deep learning algorithms: (i) DR-GCN, Eth2Vec, MODNN, SoliAudit \cite{zhang2022smart}; (i) CNN, DT, KNN, LR, RF, SVM \cite{liao2019soliaudit}; (ii) DT, NN, RF, SVM \cite{momeni2019machine}; (iii) Vanilla-RNN, LSTM, BLSTM, BLSTM-ATT \cite{qian2020towards}; (iv) KNN, RF, SVM \cite{song2019efficient}; (v) AdaBoost, KNN, RF, SVM, XGBoost \cite{wang2020contractward}. 
Also, multiple authors proposed a novel classification algorithm: the Average Stochastic Gradient Descent Weighted Dropped Long Short Term Memory \cite{gogineni2020multi}, Bytecode matching \cite{huang2021hunting}, Peculiar \cite{wu2021peculiar}, DeeSCVHunter \cite{yu2021deescvhunter}, SmartEmbed \cite{gao2020deep}, and Eth2Vec \cite{ASHIZAWA2022100101}. They usually compared the performance of their solution with baseline models. 

As far as the features fed to the models are concerned, the opcode is a popular feature \cite{zhang2022smart,liao2019soliaudit,gogineni2020multi}. Furthermore, the authors in \cite{momeni2019machine} extracted 17 features from the smart contract code: (i) features representing the execution path, and (ii) features representing heuristic guesses of the complexity of the code. In \cite{ASHIZAWA2022100101}, the authors used the same features, except for the hexadecimal address. Other features used in the papers we studied include: contract snippets \cite{qian2020towards}, bigram features from simplified opcodes \cite{song2019efficient,wang2020contractward}, bytecode \cite{huang2021hunting}, crucial data flow graph \cite{wu2021peculiar}, and vulnerability candidate slice \cite{yu2021deescvhunter}. 

Except for the study by \cite{qian2020towards}, all papers used datasets with more than 10,000 data points. Also, no authors reported a dataset spanning after 2018.

\begin{longtable}{p{3.5cm}p{1.5cm}p{1.5cm}p{3cm}p{4.5cm}}
\toprule
 Paper & Blockchain & Data-Code   & Model    & Performance\\ 
\\ 
\midrule
\cite{zhang2022smart} 
    & Ethereum
    & Y-Y
    & Classification and Deep Learning: \textbf{DR-GCN}, Eth2Vec, \textbf{MODNN}, SoliAudit
    & Accuracy, Euclidian Distance, F1 (0.9628), FN, FP, Fowlkes and Mallows Index, Loss function, Precision (0.9651), Recall (1.0), TN, TP \\
\cite{gogineni2020multi} 
    & Ethereum
    & Y-N
    & Deep Learning: \textbf{Average Stochastic Gradient Descent Weighted Dropped Long Short Term Memory}
    & Accuracy (0.913), F1 (0.952), Precision (0.944), Recall (0.979)\\
\cite{huang2021hunting} 
    & Ethereum
    & N-N
    & Classification: Bytecode matching
    & FNR (0.1429), FPR (0.0), Precision (0.9195) \\
\cite{liao2019soliaudit} 
    & Ethereum
    & N-N
    & Classification and Deep Learning: CNN, DT, KNN, \textbf{LR}, RF, SVM
    & Accuracy (0.973), F1 (0.904), Precision (0.929), Recall (0.882)\\
\cite{momeni2019machine} 
    & Ethereum
    & N-N
    & Classification: DT, NN, RF, SVM
    & Accuracy, F1, FN, FP, Precision, Recall, TN, TP \\
\cite{qian2020towards} 
    & Ethereum
    & N-Y
    & Classification and Deep Learning: BLSTM, \textbf{BLSTM-ATT}, LSTM, Vanilla-RNN
    & Accuracy (0.8847), AUC, F1 (0.8826), FPR (0.0857), Precision (0.8850), Recall, ROC Curve \\
\cite{song2019efficient} 
    & Ethereum
    & N-N
    & Classification: KNN, \textbf{RF}, SVM
    & F1 (0.98), Macro-F1 (0.9303), Micro-F1 (0.9698), ROC\\
\cite{wang2020contractward} 
    & Ethereum
    & N-N
    & Classification: AdaBoost, KNN, RF, SVM, \textbf{XGBoost}
    & F1 (0.99), Macro-F1 (0.9641), Micro-F1 (0.9798), ROC \\
\cite{wu2021peculiar} 
    & Ethereum
    & Y-Y
    & Classification: Peculiar
    & F1 (0.9210), Precision (0.9180), Recall (0.9240)\\
\cite{yu2021deescvhunter} 
    & Ethereum
    & Y-N
    & Classification: \textbf{DeeSCVHunter}
    & Accuracy (0.9302), F1 (0.8687), Precision (0.9070), Recall (0.8346)\\
\cite{gao2020deep} 
    & Ethereum
    & Y-Y
    & Deep Learning: Sequence to sequence recurrent autoencoder
    & - \\
\cite{ASHIZAWA2022100101} 
    & Ethereum
    & Y-Y
    & Classification: \textbf{Eth2Vec}, NNN, SVM
    & F1 (0.575), Precision (0.77), Recall (0.507)\\

\bottomrule
\caption{Papers Addressing Smart Contract Vulnerability Detection}
\end{longtable}
\label{tab:sc-vulnerability}

\subsection{Other} \label{sec:other}
We found three papers that do not fit the identified categories above. Specifically, \cite{huang2017behavior} addressed behavior pattern clustering. The authors used a dataset of less than 2,000 data points. Next, \cite{norvill2017automated} proposed a solution for smart contracts labeling on Ethereum, with a dataset of less than 1,000 data points. The authors used k-medoids and Affinity Propagation algorithms to obtain seven clusters. Finally, in \cite{song2022eos} , the authors applied a clustering algorithm on a dataset of less than 100,000 data points. The goal was to perform some data analysis on the EOS blockchain. 

\begin{longtable}{p{3.5cm}p{1.5cm}p{1.5cm}p{3cm}p{4.5cm}}
\toprule
 Paper & Blockchain & Data-Code   & Model    & Performance\\ 
\\ 
\midrule
\cite{huang2017behavior} 
    & Unclear
    & N-N
    & Clustering: \textbf{BPC}, BPC with Random Cluster Center Initialization
    & Precision (0.7426)\\
\cite{norvill2017automated} 
    & Ethereum
    & N-N
    & Clustering: Affinity Propagation, k-Medoids
    & Frequency distribution\\
\cite{song2022eos} 
    & EOSIO
    & Y-N
    & Clustering
    & Average Shortest Path Length, Clustering coefficient (0.448), Degree Distribution\\
\bottomrule
\caption{Papers Addressing Other Use Cases}
\end{longtable}
\label{tab:other}

\section{Discussion \& Conclusion} \label{sec:discussion}
\subsection{Major Findings}
In this work, we conducted a systematic mapping study on Machine Learning techniques applied to blockchain data. We analyzed 159 articles, selected and classified them according to various dimensions.
In order to perform this study, we followed the most widely accepted guidelines in the literature and gave an overview of the current research in the given area.
This structured methodology allowed us to extract key information and address the four main research questions. We also identified the gaps in the existing research works and recognized future research directions. 

In this study, we selected the most relevant and important publications by applying inclusion and exclusion criteria in a process consisting of 6 phases, while querying in the most widely used scientific databases in these fields i.e. Google Scholar, Springer, and ScienceDirect. The publication forums of the studies, as well as the distribution of the studies by publication type were extensively analyzed.

Followingly, the studies were classified according to four different perspectives: 
\begin{enumerate}
\item \textbf{Domain \& Use Case} with the major use cases being address classification, anomaly detection, cryptocurrency price prediction, performance prediction, smart contract vulnerability detection. The majority
of the studies we considered here addressed the problems of anomaly detection.
\item \textbf{Blockchain Platform} with the majority of studies focusing on Bitcoin and Ethereum data, while some research teams focused on a combination of blockchains.  
\item \textbf{Data} with detailed analysis on the blockchain and external data sources as well as the size of datasets examined by the different studies to draw interesting results.  It was inferred that the majority of studies were based on datasets of more than 1,000,000 data points and while some datasets were made available by the authors or were available by default, the majority of studies did not share the data they used.
\item \textbf{Machine Learning Models} applied to blockchain data such as classification, clustering, deep learning, graph learning, regression, time series analysis, or a combination of the above, with classification being the most widely used technique. Regarding the machine learning workflow followed in the different studies, some preprocessing techniques were commonly employed such as data preprocessing, feature extraction, and algorithm selection. On the subject of the specific algorithms,  Random Forest and the Support Vector Machine were the most favored ones and the most commonly utilized evaluation metrics  across studies were Accuracy, Precision, Recall, F-Score, and ROC Curve. 
\end{enumerate}

\subsection{Challenges and Future Research Directions}
As blockchains are increasingly popular, their size increases significantly depending on the number of transactions, the block size, the frequency of the block creation, and the additional data stored e.g. Ethereum smart contract events and metadata. Larger data size, which currently exceeds hundred of Gigabytes for the most popular blockchains, pose challenges in terms of storage, and computational requirements in applying machine learning to blockchain data and fully leverage the potential of machine learning algorithms in this context. 
Furthermore, to exploit the significant amount of valuable information stored in the different smart contracts and Decentralized Applications requires an additional notable effort to preprocess and often to develop solutions in order to extract it from blocks and transactions. This challenge requires further attention and it is indicative that there is an initial effort in this direction for Ethereum DApps \cite{min2022portrait} and DeFi \cite{palaiokrassas2023leveraging} with many potentials by taking also into consideration the interoperability between different blockchains and external data sources.

While some research works such as \cite{weber2019anti} \cite{farrugia2020detection} published their underlying datasets and acted as a reference point and a comparison baseline, most works are not making their dataset/code available and as a result, there is a lack of standardized evaluation frameworks and benchmarks. An additional challenge would be the constant update of such benchmarks with the latest block data and the domain's changing patterns. As an example in the domain of Anomaly Detection new malicious accounts, illicit transactions, Ponzi schemes based on smart contracts, and phishing activities constantly emerge, based on novel patterns and mechanisms and the efficient proposed algorithms should be able to identify the most recent malicious activities. This lack of data sharing is a common problem in academia, and many reasons explain this issue, such as insufficient time, lack of funding, lack of skills, or authors' attitudes. Various mechanisms could be put in place in order to promote or facilitate data sharing, including: explicit data sharing policies from journals, support from authors' institutions, better infrastructure/data repositories, or incentives for sharing \cite{tenopir2011data,kim2015understanding,zenk2018factors,perrier2020views}. 

Finally, in relation to the points raised above, blockchain is a fast evolving technology: new blockchains, new protocols, new DApps, new use cases, and new interactions with other technologies (blockchain and IoT, blockchain and AI) emerge constantly. Being able to store and process the growing volumes of data is crucial, as well as being able to compare past phenomena with the current ones. The latter would be made possible with proper data sharing. 

\subsection{Conclusion}
In summary, this study provided detailed insights into the research works on applying machine learning to blockchain data. 
As the adoption of blockchains continues to grow, new technical solutions are integrated and new services emerge. Future research needs to take into consideration such aspects in order to better understand the research landscape. Novel research is needed to propose standardized frameworks across the different use cases analyzed in the context of this work. As significant growth in machine learning research is noticed and the proposal of innovative ML algorithms is observed, future research is also required to explore their tailoring, customization and potential for the unique features of blockchain problems and data.

\section*{Acknowledgment}
 This work was supported by J.P.Morgan Faculty Research Awards 2022 and NSF 1932220.


\bibliography{references}

\end{document}